\documentclass[preprints,article,accept,moreauthors,pdftex]{Definitions/mdpi} 

\firstpage{1} 
\pubvolume{1}
\issuenum{1}
\articlenumber{0}
\pubyear{2022}
\copyrightyear{2022}
\externaleditor{Academic Editor: Firstname Lastname} 
\datereceived{13 May 2022} 
\dateaccepted{8 June 2022} 
\datepublished{} 
\hreflink{https://doi.org/} 

\Title{Onset of Electron Captures and Shallow Heating in Magnetars}

\TitleCitation{Onset of Electron Captures and Shallow Heating in Magnetars}

\Author{Nicolas Chamel $^{1,}$*\orcidA{} 
	and Anthea Francesca Fantina $^{2,1}$\orcidB{}} 

\AuthorNames{Nicolas Chamel and Anthea Francesca Fantina}

\AuthorCitation{Chamel, N.; 
	Fantina, A.F.}

\address{
	$^{1}$ \quad Institute of Astronomy and Astrophysics, Universit\'e Libre de Bruxelles, CP 226, Boulevard du Triomphe, B-1050 Brussels, Belgium; 
	nicolas.chamel@ulb.be
	\\
	$^{2}$ \quad Grand Acc\'el\'erateur National d'Ions Lourds (GANIL), CEA/DRF-CNRS/IN2P3, Boulevard Henri Becquerel, 14076 Caen, France; 
	anthea.fantina@ganil.fr}

\corres{Correspondence: nicolas.chamel@ulb.be} 

\abstract{The loss of magnetic pressure accompanying the decay of the magnetic field in a magnetar may trigger exothermic electron captures by nuclei in the shallow layers of the stellar crust. Very accurate analytical formulas are obtained for the threshold density and pressure, as well as for the maximum amount of heat that can be possibly released, taking into account the Landau–Rabi quantization of electron motion. These formulas are valid for arbitrary magnetic field strengths, from the weakly quantizing regime to the most extreme situation in which electrons are all confined to the lowest level. Numerical results are also presented based on experimental nuclear data supplemented with predictions from the Brussels-Montreal model HFB-24. This same nuclear model has been already employed to calculate the equation of state in all regions of magnetars. 
}

\keyword{\textls[-20]{neutron star; magnetar; outburst; magnetic field; shallow heating; cooling; electron capture}} 

\def\apj{Astrophys. J.}  
\def\apjl{Astrophys. J. Lett.} 
\def\prc{Phys.~Rev.~C} 
\def\prd{Phys.~Rev.~D} 
\def\pre{Phys.~Rev.~E} 
\def\aap{Astron. Astrophys.}   
\def\nat{Nature}
\def\mnras{Mon. Not. R. Astron. Soc.}             
\def\apjs{Astrophys. J. Suppl.}  

\begin{document}

\section{Introduction}

Soft gamma-ray repeaters and anomalous X-ray pulsars are two facets of a very active subclass of neutron stars, called magnetars, exhibiting outbursts and less frequently giant flares that release huge amounts of energy up to $\sim$$10^{46}$~erg within a second (see e.g.,~\cite{esposito2021} for a recent review). These phenomena are thought to be powered by internal magnetic fields exceeding $10^{14}$--$10^{15}$~G~\cite{duncan1992}. At~the date of this writing, 24 such objects have been discovered and six more candidates remain to be confirmed according to the McGill Online Magnetar Catalog~\cite{olausen2014}. Their persistent X-ray luminosity $\sim$$10^{33}$--$10^{35}$ erg/s, which is well in excess of their rotational energy and which implies a
higher surface temperature
than in weakly magnetized neutron stars of the same age~\cite{vigano2013}, provides further evidence for extreme magnetic fields~\cite{beloborodov2016}. It is widely thought that heat is generated by the deformations of the crust beyond the elastic limit due to magnetic stresses 
(see, e.g.,~\cite{degrandis2020}). This mechanism is most effective in the inner region of the crust, where 
crystallization first occurs~\cite{fantina2020,carreau2020}. However, it has been demonstrated that heat sources should be located in the shallow region of the crust to avoid excessive neutrino losses~\cite{kaminker2006,kaminker2009}. Alternatively, the~magnetic energy in the outer crust may be dissipated into heat through electron captures by nuclei triggered by the magnetic field evolution~\cite{cooper2010}. This mechanism is analogous to crustal heating in accreting neutron stars~\cite{haensel1990}, the~matter compression being induced here by the loss of magnetic support rather than accretion from a stellar~companion. 

We have recently estimated the maximum amount of heat that could be possibly released by electron captures and the location of the heat sources taking into account Landau--Rabi quantization of electron motion induced by the magnetic field~\cite{chamel2021}. For~simplicity, we focused on the strongly quantizing regime in which only the lowest Landau--Rabi level is occupied, thus allowing for a simple analytical treatment. Results are extended here to arbitrary magnetic fields. We demonstrate that the weakly quantizing regime is also amenable to accurate analytical approximations. 
Results are presented based on experimental nuclear data supplemented with the Brussels-Montreal atomic mass table HFB-24~\cite{goriely2013}. The~underlying nuclear energy-density functional BSk24 has been already applied to construct unified equations of state for both unmagnetized neutron stars~\cite{pearson2018,pearson2020,pearson2022} and magnetars~\cite{mutafchieva2019}. 

The paper is organized as follows. In~Section~\ref{sec:eos}, we present the equation of state of the outer crust of a magnetar and the approximations we made. 
In Section~\ref{sec:initial-composition}, we give the equations to determine the initial composition of the outer crust and the boundaries delimiting different layers. Our analytical treatment of the heating 
from electron captures is described in Section~\ref{sec:heating}. Numerical results including detailed error estimates are presented and discussed in Section~\ref{sec:results}.

\section{Equation of State of Magnetar~Crusts}
\label{sec:eos}

In the following, we shall consider the crustal region at densities above the ionization threshold and below the neutron-drip point. We assume that each crustal layer is made of fully ionized atomic nuclei $(A,Z)$ with proton number $Z$ and mass number $A$ embedded in a relativistic electron~gas.

\subsection{Main~Equations}

Whereas nuclei with number density $n_N$ exert a negligible pressure $P_N\approx 0$, they contribute to the mass-energy density
\begin{equation}
	\label{eq:rho}
	\mathcal{E}_N=n_N M^\prime(A,Z,B_\star)c^2\, ,
\end{equation}
where $M^\prime(A,Z,B_\star)$ denotes the ion mass including the rest mass of $Z$ electrons. In~principle, $M^\prime(A,Z,B_\star)$ may also depend on the magnetic field, which will be conveniently measured in terms of the dimensionless ratio $B_\star\equiv B/B_{\rm rel}$ with
\begin{equation}
	\label{eq:Bcrit}
	B_{\rm rel}=\frac{m_e^2 c^3}{e \hbar}\approx 4.41\times 10^{13}\, \rm G\, , 
\end{equation}
where $m_e$ is the electron mass, $c$ is the speed of light, $\hbar$ is the Planck--Dirac constant and $e$ is the elementary electric~charge.

To a very good approximation, electrons can be treated as an ideal Fermi gas. In~the presence of a  magnetic field, the~electron motion perpendicular to the field is quantized into Landau--Rabi  
levels~\cite{rabi1928,landau1930}. The~observed surface magnetic field on a magnetar is typically $B_s\sim 10^{14}$--$10^{15}$~G~\cite{olausen2014,tiengo2013,hongjun2014}. The~internal magnetic field $B$ is expected to be even stronger and could potentially reach $10^{17}-10^{18}$~G 
(see, e.g.,~\cite{uryu2019}). In~our previous study~\cite{chamel2021}, we assumed for simplicity that the magnetic field is strongly quantizing, meaning that electrons remain all confined to the lowest level throughout the outer crust, thus requiring  $B\gtrsim 5.72\times 10^{16}$~G~\cite{chamel2015b}. 
Even if weaker fields are considered, quantization effects are not expected to be completely washed out by thermal effects. Indeed, 
the temperatures $T\sim 10^8$--$10^9$~K prevailing in a magnetar for which $B_\star\gg 1$ 
(see e.g.,~\cite{kaminker2009}) are much lower than the characteristic temperature
\begin{equation}\label{eq:TB}
	T_B=\frac{m_e c^2}{k_B} B_\star\approx 5.93\times 10^9 B_\star~\rm K\, , 
\end{equation}
where $k_B$ denotes Boltzmann's constant. 
Strictly speaking, Equation~\eqref{eq:TB} is only relevant in the strongly quantizing regime. If~several Landau--Rabi levels are populated, the~characteristic temperature is reduced but only by a factor of a few at most at the bottom of the outer crust (see Chap. 4 in~\cite{haensel2007}). 
Neglecting the small electron anomalous magnetic moment and ignoring thermal effects, 
the electron energy density (with the rest-mass excluded) and electron pressure are given by
\begin{equation}
	\label{eq:energy-electron}
	\mathcal{E}_e=\frac{B_\star m_e c^2}{(2 \pi)^2 \lambda_e^3} \sum_{\nu=0}^{\nu_{\rm max}}g_\nu(1+2\nu B_\star) \psi_+ \biggl[\frac{x_e(\nu)}{\sqrt{1+2\nu B_\star}}\biggr]-n_e m_e c^2\, ,
\end{equation}
\begin{equation}
	\label{eq:pressure-electron}
	P_e=\frac{B_\star m_e c^2}{(2 \pi)^2 \lambda_e^3} \sum_{\nu=0}^{\nu_{\rm max}}g_\nu(1+2\nu B_\star) \psi_- \biggl[\frac{x_e(\nu)}{\sqrt{1+2\nu B_\star}}\biggr]\, ,
\end{equation}
respectively, where we have introduced the electron Compton wavelength $\lambda_e=\hbar/(m_e c)$, $g_\nu=1$ for $\nu=0$ and $g_\nu=2$ for $\nu\geq 1$,
\begin{equation}
	\label{eq:psi}
	\psi_\pm(x)=x\sqrt{1+x^2}\pm\ln(x+\sqrt{1+x^2})\, ,
\end{equation}
\begin{equation}
	\label{eq:xe}
	x_e(\nu) =\sqrt{\gamma_e^2 -1-2 \nu B_\star}\, ,
\end{equation}
and $\nu_\textrm{max}$ is fixed by the electron number density $n_e$ given by
\begin{equation}
	\label{eq:ne}
	n_e =\frac{2 B_\star}{(2 \pi)^2 \lambda_e^3} \sum_{\nu=0}^{\nu_{\rm  max}} g_\nu x_e(\nu)\, .
\end{equation}
Here $\gamma_e$ denotes the electron Fermi energy in units of $m_e c^2$. 
The index $\nu_\text{max}$ is the highest integer for which $\gamma_e^2-1-2\nu_{\rm  max}B_\star\geq 0$, i.e.
\begin{equation}
	\nu_\text{max}=\left[\frac{\gamma_e^2-1}{2 B_\star}\right]\, , 
\end{equation}
where $[.]$ denotes the integer part. The~mean baryon number density follows from the requirement of electric charge neutrality
\begin{equation}
	\bar n=\frac{A}{Z}n_e= A n_N\, .
\end{equation}

The main correction to the ideal electron Fermi gas arises from the electron-ion interactions. 
According to the Bohr-van Leeuwen theorem~\cite{vanvleck1932}, the~electrostatic corrections to the energy density and to the pressure are 
independent of the magnetic field apart from a negligibly small contribution due to quantum zero-point motion 
of ions about their equilibrium position~\cite{baiko2009}. For~pointlike ions embedded in a uniform electron gas,  
the corresponding energy density is given by (see e.g.,~Chap. 2 of~\cite{haensel2007})
\begin{equation}\label{eq:EL}
	\mathcal{E}_L=C_M  \left(\frac{4\pi}{3}\right)^{1/3} e^2 n_e^{4/3} Z^{2/3}\, ,
\end{equation}
where $C_M$ is the Madelung constant. 
The contribution to the pressure is thus given by
\begin{equation}\label{eq:PL}
	P_L=n_e^2 \frac{d(\mathcal{E}_L/n_e)}{dn_e}=\frac{\mathcal{E}_L}{3}\, . 
\end{equation}

The pressure of the Coulomb plasma finally reads $P=P_e+P_L$, whereas the energy density is given by $\mathcal{E}=\mathcal{E}_N+\mathcal{E}_e+\mathcal{E}_L$. 

For ions arranged in a body-centered cubic lattice, the~Madelung constant is given by $C_M=-0.895929255682$~\cite{baiko2001}. However, the~electron-ion plasma may not necessarily be in a solid state, especially in the shallow layers, which are the main focus of this work. The~crystallization temperature can be estimated as~\cite{fantina2020}:
\begin{equation}\label{eq:Tm}
	T_m\approx 1.3\times 10^5 Z^2 \left(\frac{175}{\Gamma_m}\right) \left(\frac{\rho_6}{A}\right)^{1/3}~{\rm K}\, ,
\end{equation}
where $\rho_6$ is the density in units of $10^6$~g~cm$^{-3}$, and~$\Gamma_m$ is the Coulomb 
coupling parameter at melting. In~the absence of magnetic field, $\Gamma_m\approx 175$ and 
$T_m$ is typically of order $10^9$~K. The~presence of a magnetic field tends to lower 
$\Gamma_m$, thus increasing $T_m$~\cite{potekhin2013}. In~any case, the~Madelung constant in the 
liquid phase remains very close to that of the solid phase. In~ the following, we will adopt the 
Wigner--Seitz estimate $C_M=-9/10$ for the Madelung constant~\cite{salpeter1954}. Thermal effects on 
thermodynamic quantities are small and will be~neglected.

\subsection{Weakly Quantizing Magnetic~Field}
\label{subsec:weakly-quantizing}

\textls[-10]{The magnetic field is weakly quantizing if many Landau--Rabi levels are filled: \mbox{$\nu_{\rm max}\gg 1$.} }
Using the expansions (41) obtained in~\cite{dib2001} for the electron density $n_e$ 
leads to the following estimate for the mean baryon number density:
\begin{equation}\label{eq:rho-mag-weak}
	\bar n\approx \frac{A}{2\pi^2 Z \lambda_e^3} \bigg[ \frac{2}{3}\left(\gamma_e^2-1\right)^{3/2}+(2B_\star)^{3/2}\zeta\left(\frac{-1}{2},\left\{\frac{\gamma_e^2-1}{2B_\star}\right\}\right)+\frac{B_\star^2}{6\sqrt{\gamma_e^2-1}}\biggr]\, , 
\end{equation} 
where $\zeta(z,q)$ is the Hurwitz zeta function defined by
\begin{equation}
	\zeta(z,q)=\sum_{\nu=0}^{+\infty} \frac{1}{(\nu+q)^z}
\end{equation} 
for $\Re(z)>1$ and by analytic continuation to other $z\neq 1$ (excluding poles $\nu+q=0$).
The first term in Equation~\eqref{eq:rho-mag-weak} represents the mean baryon number density in the absence of magnetic field. The~second term accounts for quantum oscillations due to the filling of Landau--Rabi levels, while the last term is a higher-order magnetic~correction. 

The expression for the associated expansion of the pressure is more involved. 
In the notations of~\cite{dib2001}, the~electron contribution $P_e$ to the pressure can be directly obtained from the grand potential density by $P_e=-\omega_0^{(\rm mon)}-\omega_0^{(\rm osc)}$. Using Equations~(41), (43) and (44) of \cite{dib2001} yields ($\alpha=e^2/(\hbar c)$ is the fine-structure constant):
\begin{footnotesize}
	\begin{align}\label{eq:electron-pressure-mag-weak}
		P_e & \approx \frac{m_e c^2}{4\pi^2 \lambda_e^3}\biggl\{\frac{1}{2}\left(1-2B_\star+\frac{2B_\star^2}{3}\right)\log\left(\frac{\gamma_e+\sqrt{2B_\star+\gamma_e^2-1}}{1+\sqrt{2B_\star}}\right)\nonumber \\ &-\frac{1}{2}\left(\gamma_e\sqrt{2B_\star+\gamma_e^2-1}-\sqrt{2B_\star}\right)+\frac{1}{3}\left(\gamma_e\sqrt{2B_\star+\gamma_e^2-1}^3-\sqrt{2B_\star}^3\right)\nonumber \\ 
		&+B_\star \left({\rm arccosh}~\gamma_e - \gamma_e\sqrt{\gamma_e^2-1}\right)-(2B_\star)^{5/2} \int_0^{+\infty}\frac{\tilde{\zeta}_3(-1/2,q+1)}{\sqrt{1+2B_\star q}} dq \nonumber \\
		&+ \frac{2}{3} \frac{(2 B_\star)^{5/2}}{\gamma_e}\zeta\left(\frac{-3}{2},\left\{\frac{\gamma_e^2-1}{2B_\star}\right\}\right)  
		+\frac{2}{15}\frac{(2B_\star)^{7/2}}{\gamma_e^3}\zeta\left(\frac{-5}{2},\left\{\frac{\gamma_e^2-1}{2B_\star}\right\}\right)  \nonumber \\
		&+\frac{1}{240}\left(\frac{B_\star}{\gamma_e}\right)^4 +4B_\star^2 \int_0^1 \zeta\left(\frac{-1}{2},q\right)\zeta\left(\frac{1}{2},q+\frac{1}{2B_\star}\right)dq \biggr\}
		\, ,
	\end{align}
\end{footnotesize}
with
\begin{equation}
	\tilde{\zeta}_3(z,q)=\zeta(z,q)-\frac{1}{z-1}q^{-z+1}-\frac{1}{2}q^{-z}-\frac{z}{12}q^{-z-1}\, .
\end{equation}

\textls[20]{The total pressure is found by adding the electrostatic correction~\eqref{eq:PL} using the expansion for the} \mbox{electron~density:}
\begin{footnotesize}
	\begin{align}\label{eq:pressure-mag-weak}
		P & \approx \frac{m_e c^2}{4\pi^2 \lambda_e^3}\biggl\{\frac{1}{2}\left(1-2B_\star+\frac{2B_\star^2}{3}\right)\log\left(\frac{\gamma_e+\sqrt{2B_\star+\gamma_e^2-1}}{1+\sqrt{2B_\star}}\right)\nonumber \\ &-\frac{1}{2}\left(\gamma_e\sqrt{2B_\star+\gamma_e^2-1}-\sqrt{2B_\star}\right)+\frac{1}{3}\left(\gamma_e\sqrt{2B_\star+\gamma_e^2-1}^3-\sqrt{2B_\star}^3\right)\nonumber \\ 
		&+B_\star \left({\rm arccosh}~\gamma_e - \gamma_e\sqrt{\gamma_e^2-1}\right)-(2B_\star)^{5/2} \int_0^{+\infty}\frac{\tilde{\zeta}_3(-1/2,q+1)}{\sqrt{1+2B_\star q}} dq \nonumber \\
		&+ \frac{2}{3} \frac{(2 B_\star)^{5/2}}{\gamma_e}\zeta\left(\frac{-3}{2},\left\{\frac{\gamma_e^2-1}{2B_\star}\right\}\right)  
		+\frac{2}{15}\frac{(2B_\star)^{7/2}}{\gamma_e^3}\zeta\left(\frac{-5}{2},\left\{\frac{\gamma_e^2-1}{2B_\star}\right\}\right)  \nonumber \\
		&+\frac{1}{240}\left(\frac{B_\star}{\gamma_e}\right)^4 +4B_\star^2 \int_0^1 \zeta\left(\frac{-1}{2},q\right)\zeta\left(\frac{1}{2},q+\frac{1}{2B_\star}\right)dq \nonumber \\ 
		&+\frac{2}{3}\left(\frac{2}{3\pi}\right)^{1/3} C_M \alpha Z^{2/3}\biggl[  \frac{2}{3}\left(\gamma_e^2-1\right)^{3/2} \nonumber \\ 
		&+(2B_\star)^{3/2}\zeta\left(\frac{-1}{2},\left\{\frac{\gamma_e^2-1}{2B_\star}\right\}\right)+\frac{B_\star^2}{6\sqrt{\gamma_e^2-1}}\biggr]^{4/3} \biggr\}
		\, .
	\end{align}
\end{footnotesize}

In the absence of magnetic field $B_\star=0$ (corresponding to the limit $\nu_{\rm  max}\rightarrow +\infty$), the~mean baryon number density and the pressure reduce, respectively, to
\begin{eqnarray}\label{eq:rho-nomag}
	\bar n  = \frac{A}{Z}  \frac{x_r^3 }{3 \pi^2 \lambda_e^3}\, , 
\end{eqnarray}\vspace{-9pt}
\begin{eqnarray}
	\label{eq:pressure-nomag}
	P&=&\frac{m_e c^2}{8 \pi^2 \lambda_e^3} \biggl[x_r\left(\frac{2}{3}x_r^2-1\right)\sqrt{1+x_r^2}+\ln(x_r+\sqrt{1+x_r^2})\biggr]\nonumber\\
	&&+\frac{C_M \alpha}{3}\left(\frac{4}{243\pi^7}\right)^{1/3} x_r^4 \frac{m_e c^2}{\lambda_e^3}Z^{2/3} \, .
\end{eqnarray}
Here, $x_r$ denotes the relativity parameter defined by $x_r=\lambda_e k_e$ and $k_e=(3\pi^2 n_e)^{1/3}$ is the electron Fermi wave number. The~Fermi energy is then given by
\begin{equation}\label{eq:gammae-nomag}
	\gamma_e =  \sqrt{1+x_r^2}\, .
\end{equation}

\begin{figure}[H]
	
	\includegraphics[width=10.5cm]{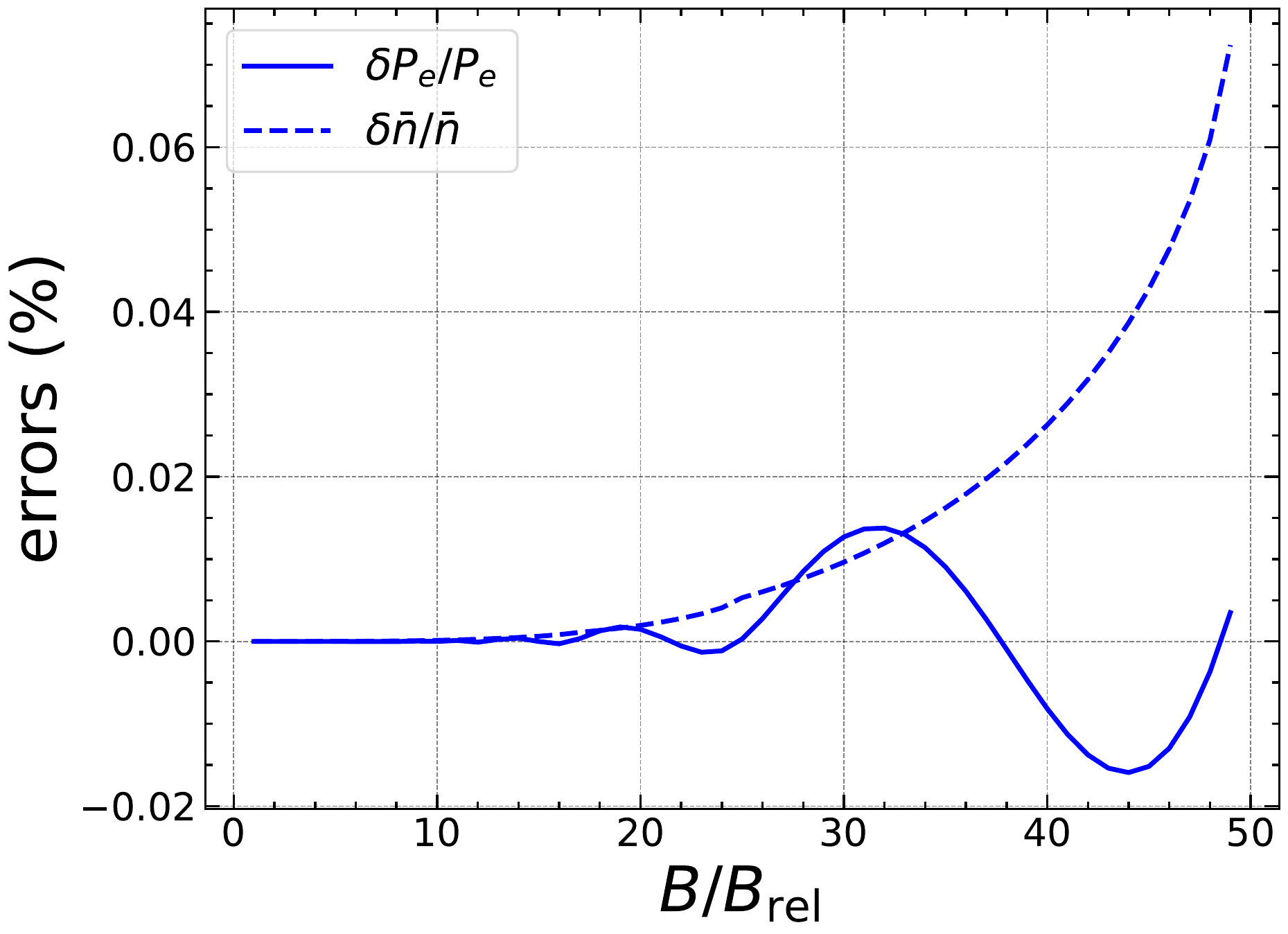}
	\caption{Relative errors (in \%) of the approximate analytical expansions \eqref{eq:rho-mag-weak} and \eqref{eq:electron-pressure-mag-weak} for the baryon number density $\bar n$ (dashed line) and electron pressure $P_e$ (solid line), respectively, as~a function of the magnetic field strength $B_\star=B/B_{\rm rel}$ with $\gamma_e=10$. The~errors are obtained by taking the difference between the approximate and exact results and dividing by the exact result. All electrons are confined to the lowest Landau--Rabi level at $B_\star=(\gamma_e^2-1)/2=49.5$.
	}

	\label{fig1}
\end{figure}

\begin{figure}[H]
	
	\includegraphics[width=10.5cm]{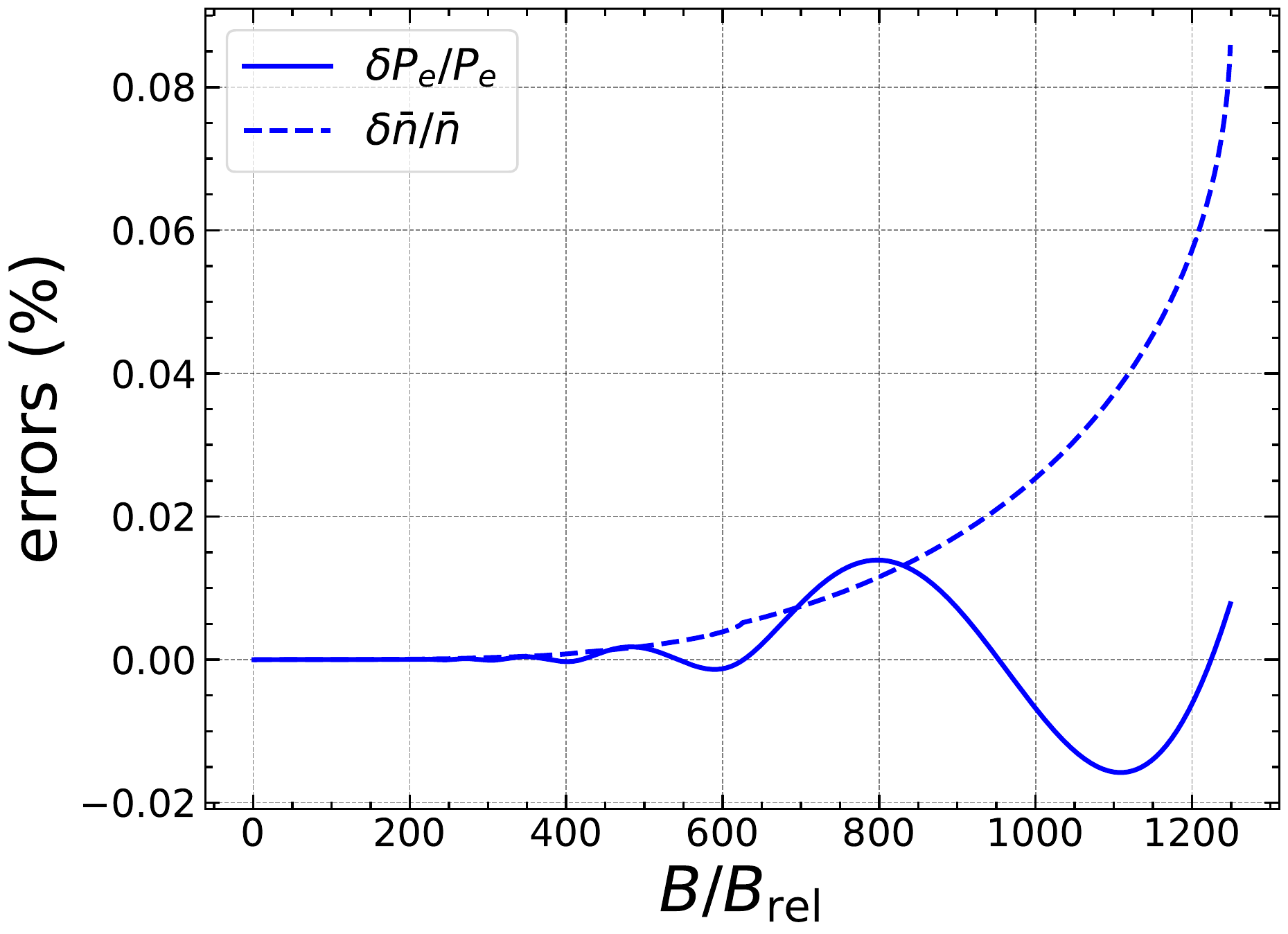}
	\caption{Same as Figure~\ref{fig1} for $\gamma_e=50$. All electrons are confined to the lowest Landau--Rabi level at $B_\star=(\gamma_e^2-1)/2=1249.5$.
	}
	\label{fig1b}
\end{figure}

As shown in Figures~\ref{fig1} and \ref{fig1b} for two representative values $\gamma_e=10$ (shallow region of the outer crust) and $\gamma_e=50$ (bottom of the outer crust), respectively, the~expansions \eqref{eq:rho-mag-weak} and \eqref{eq:electron-pressure-mag-weak} are surprisingly precise throughout the outer crust. In~the limit of vanishingly small magnetic field ($\nu_{\rm max}\rightarrow +\infty$), 
Equations~\eqref{eq:rho-mag-weak} and \eqref{eq:electron-pressure-mag-weak} converge toward the exact results, \eqref{eq:rho-nomag} and \eqref{eq:pressure-nomag}, respectively.  Although~the errors increase with the magnetic 
field as expected, they 
remain very small in the intermediate regime $\nu_{\rm max}\sim 1$ for which the field is no longer weakly quantizing. 
When electrons start to be all confined to the lowest Landau--Rabi level, i.e.,~when $\nu_{\rm max}=0$, the~error on 
$\bar n$ amounts to 0.1\% only. The~approximate formula for the pressure is found to be more reliable, with~errors 
not exceeding 0.02\% and~fluctuating.

The expansions \eqref{eq:rho-mag-weak} and \eqref{eq:electron-pressure-mag-weak}
can thus be confidently applied for arbitrary magnetic field strengths, from~$B_\star=0$ ($\nu_{\rm max}\rightarrow+\infty$) up to the threshold magnetic field \mbox{$B_\star=(\gamma_e^2-1)/2$} at the onset of the strongly quantizing regime ($\nu_{\rm max}=0$) discussed in the next~subsection. 

\subsection{Strongly Quantizing Magnetic~Field}

The magnetic field is strongly quantizing if all electrons lie in the lowest Landau-Rabi level. Setting $\nu_{\rm max}=0$ in Equations~\eqref{eq:pressure-electron} and \eqref{eq:ne} leads to the following expression for the pressure and mean baryon number density:
\begin{align}\label{eq:pressure-mag-strong}
	P &=\frac{B_\star m_e c^2 }{4 \pi^2 \lambda_e^3 }\biggl[\gamma_e\sqrt{\gamma_e^2-1}-\ln\left(\sqrt{\gamma_e^2-1}+\gamma_e\right) \nonumber \\ 
	&+\frac{C_M \alpha }{3}\left(\frac{16 B_\star Z^2 }{3\pi}\right)^{1/3} \left(\gamma_e^2-1\right)^{2/3} \biggr] \, ,
\end{align}
\begin{equation}\label{eq:rho-mag-strong}
	\bar n  = \frac{B_\star}{2\pi^2 \lambda_e^3} \frac{A}{Z}   \sqrt{\gamma_e^2-1} \, . 
\end{equation}

These expressions hold if
\begin{equation}\label{eq:gammae-mag-strong-cond}
	\gamma_e < \sqrt{1+2B_\star}	\, , 
\end{equation} 
or equivalently if
\begin{equation}\label{eq:ne-mag-strong-cond}
	n_e<\frac{B_{\star}^{3/2}}{\sqrt{2}\pi^2\lambda_e^3}  \, , 
\end{equation}
using Equations~\eqref{eq:xe} and \eqref{eq:ne}.

\section{Initial Composition of Magnetar~Crusts}
\label{sec:initial-composition}

Assuming the crust is initially in a full thermodynamic equilibrium in the presence of some magnetic 
field, the~composition is found by minimizing the Gibbs free energy per nucleon, which coincides with the baryon chemical potential (see, e.g.,~Appendix A in~\cite{fantina2015}):
\begin{equation}
	\label{eq:gibbs}
	\mu=\frac{\mathcal{E}+P}{\bar n}=\frac{M^\prime(A,Z,B_\star)c^2}{A}+\frac{Z}{A}m_e c^2\biggl[ \gamma_e -1 +\frac{4}{3} C_M\left(\frac{4\pi}{3}\right)^{1/3} \alpha \lambda_e n_e^{1/3} Z^{2/3}\biggr]\, .
\end{equation} 
This minimization can be performed very efficiently following the iterative approach proposed in~\cite{chamel2020,chamelstoyanov2020}. Freely available computer codes in the limiting cases $\nu_{\rm max}\rightarrow +\infty$ and $\nu_{\rm max}=0$ can be found in~\cite{chamel2020zndo,chamelstoyanov2020zndo}. 

\subsection{Interface between Adjacent Crustal~Layers}

The pressure $P_{1\rightarrow2}$ associated with the transition from a crustal layer made of nuclei ($A_1$, $Z_1$) to 
a denser layer made of nuclei ($A_2$, $Z_2$) is determined by the equilibrium condition
\begin{equation}
	\label{eq:crust-equilibrium}
	\mu(A_1,Z_1,P_{1\rightarrow2})=\mu(A_2,Z_2,P_{1\rightarrow2})\, ,
\end{equation}
which can be approximately written as~\cite{chamel2020}
\begin{equation}\label{eq:crust-condition}
	\gamma_e +  C_M\, \alpha \lambda_e F(Z_1,A_1 ; Z_2, A_2) \left(\frac{4\pi n_e}{3}\right)^{1/3} = \gamma_e^{1\rightarrow 2}\, ,
\end{equation}
\begin{equation}\label{eq:F12}
	F(Z_1, A_1 ; Z_2, A_2)\equiv \left(\frac{4}{3}\frac{Z_{1}^{2/3}Z_1}{A_1} - \frac{1}{3}\frac{Z_{1}^{2/3}Z_2}{A_2} -\frac{Z_{2}^{2/3}Z_2}{A_2}\right)
	\left(\frac{Z_1}{A_1}-\frac{Z_2}{A_2}\right)^{-1} \, ,
\end{equation}
\begin{equation}\label{eq:mue12}
	\gamma_e^{1\rightarrow 2}\equiv \biggl[\frac{M^\prime(A_2,Z_2,B_\star)}{ A_2 m_e}-\frac{M^\prime(A_1,Z_1,B_\star)}{A_1 m_e}\biggr]\left(\frac{Z_1}{A_1}-\frac{Z_2}{A_2}\right)^{-1} +  1\, .
\end{equation}

The bottom of the outer crust is defined by the depth at which neutrons start to drip out of nuclei. The~corresponding electron Fermi energy obeys an equation similar to Equation~\eqref{eq:crust-condition}, the~function $F(Z_1, A_1 ; Z_2, A_2)$ being replaced by $(4/3) Z^{2/3}$
and $\gamma_e^{1\rightarrow 2}$ by~\cite{chamel2015b}
\begin{equation}\label{eq:muedrip}
	\gamma_e^{\rm drip}\equiv \frac{-M^\prime(A,Z,B_\star)c^2+A m_n c^2}{Z m_e c^2} +1 \, ,
\end{equation}
where $m_n$ is the neutron~mass. 

The threshold condition~\eqref{eq:crust-condition} takes formally the same form with and without 
magnetic fields. However, the~solutions do depend on $B_\star$ through the relation between $\gamma_e$ and $n_e$, and~potentially also through the ion~masses. 

\subsection{No Magnetic~Field}

In the absence of magnetic field $B_\star=0$, the~solution of Equation~\eqref{eq:crust-condition} reads~\cite{chamel2020}
\begin{eqnarray}\label{eq:xr-threshold}
	x_r&=&\gamma_e^{1\rightarrow 2} \Biggl\{\sqrt{1-\Biggl[1-\tilde F(Z_1, A_1;Z_2,A_2)^2\Biggr]/(\gamma_e^{1\rightarrow2})^{2}}-\tilde F(Z_1, A_1; Z_2,A_2)\Biggr\}\nonumber \\
	&& \times \Biggl[1-\tilde F (Z_1, A_1; Z_2,A_2)^2\Biggr]^{-1}\, , 
\end{eqnarray}
with
\begin{equation}
	\tilde F(Z_1,A_1;Z_2,A_2) \equiv \left(\frac{4}{9\pi}\right)^{1/3} C_M \alpha F(Z_1,A_1;Z_2,A_2)\, .
\end{equation}
This solution exists only if $\tilde F(Z_1,A_1;Z_2,A_2)\geq -1$. 

The mean baryon number density $\bar n_1^{\rm max}$ up to which nuclei $(A_1,Z_1)$ are present and the transition pressure $P_{1\rightarrow 2}$ are then given by Equations~\eqref{eq:rho-nomag} and \eqref{eq:pressure-nomag} respectively. 

\subsection{Strongly Quantizing Magnetic~Field}
\label{sec:initial-composition-strongB}

The solution of Equation~\eqref{eq:crust-condition} was also found in the limit of a strongly quantizing magnetic field~\cite{chamelstoyanov2020}. Introducing
\begin{equation}
	\bar F(Z_1,A_1;Z_2,A_2; B_\star)\equiv \frac{1}{3} C_M \alpha F(Z_1,A_1;Z_2,A_2)\left(\frac{2B_\star}{3\pi}\right)^{1/3}\, ,
\end{equation}
\begin{equation}
	\upsilon\equiv \frac{\gamma_e^{1\rightarrow 2}}{2 |\bar F(Z_1,A_1;Z_2,A_2; B_\star)|^{3/2}}\, ,
\end{equation}
the electron Fermi energy at the crustal interface is given by the following formulas: 

\begin{itemize}
	\item $\gamma_e^{1\rightarrow 2}>0$ and $\bar F(Z_1,A_1;Z_2,A_2; B_\star)>0$
	\begin{equation}
		\gamma_e=8\bar F(Z_1,A_1;Z_2,A_2; B_\star)^{3/2}\, {\rm sinh}^3 \left(\frac{1}{3}{\rm arcsinh~} \upsilon\right)\, ,
	\end{equation}
	\item $\gamma_e^{1\rightarrow 2}>0$ and $\bar F(Z_1,A_1;Z_2,A_2; B_\star)<0$
	\begin{equation}
		\gamma_e=\begin{cases}
			8|\bar F(Z_1,A_1;Z_2,A_2; B_\star)|^{3/2}\, {\rm cosh}^3\left(\frac{1}{3}{\rm arccosh~} \upsilon\right) & \text{if} \ \upsilon\geq 1\, ,\\
			8|\bar F(Z_1,A_1;Z_2,A_2; B_\star)|^{3/2}\, \cos^3 \left( \frac{1}{3}\arccos \upsilon \right)& \text{if} \ 0\leq \upsilon< 1\, . 
		\end{cases}
	\end{equation}
	\item $\gamma_e^{1\rightarrow 2}<0$ and $\bar F(Z_1,A_1;Z_2,A_2; B_\star)<0$
	\begin{equation} 
		\gamma_e=8|\bar F(Z_1,A_1;Z_2,A_2; B_\star)|^{3/2}\, \cos^3\theta_k \ \ \text{if} \ -1 < \upsilon\leq 0\, , 
	\end{equation}
	\begin{equation}
		\theta_k\equiv \frac{1}{3}\arccos \upsilon + \frac{2\pi k}{3} \ \ \text{and} \ k=0,2 \, .
	\end{equation}
\end{itemize}

The mean baryon number density $\bar n_1^{\rm max}$ up to which nuclei $(A_1,Z_1)$ are present and the transition pressure $P_{1\rightarrow 2}$ are then given by Equations~\eqref{eq:rho-mag-strong} and \eqref{eq:pressure-mag-strong}, respectively. In~the case $\gamma_e^{1\rightarrow 2}<0$, the~physically admissible solution among $k=0$ and $k=2$ is the one yielding the lowest transition pressure $P_{1\rightarrow 2}$ satisfying the conditions $\gamma_e> 1$ and $\bar n_2^{\rm min} \geq \bar n_1^{\rm max}$. 

Let us recall that these solutions are only valid under the assumption $\nu_{\rm max}=0$, which translates into a lower bound for the magnetic field $B_\star \geq B^{1\rightarrow 2}_{\star}$. 
To find $B^{1\rightarrow 2}_{\star}$, we substitute Equations~\eqref{eq:gammae-mag-strong-cond} and \eqref{eq:ne-mag-strong-cond} in  Equation~\eqref{eq:crust-condition}. This leads to
\begin{equation}\label{eq:Bstar12}
	B^{1\rightarrow 2}_{\star}=\frac{(\gamma_e^{1\rightarrow 2})^2}{2}\biggl[1+\frac{C_M\alpha}{(3\pi)^{1/3}} F(Z_1,A_1;Z_2,A_2)
	\biggr]^{-2}\, .
\end{equation}
We have assumed $B^{1\rightarrow 2}_{\star}\gg 1$ so that $\gamma_e \approx \sqrt{2 B^{1\rightarrow 2}_\star}$.

\subsection{Intermediate Magnetic~Fields}

Approximate analytical solutions can also be found in the intermediate regime. 
Remarking that the magnetic field enters explicitly in  Equation~\eqref{eq:crust-condition} only through the small electrostatic correction, the~threshold electron Fermi energy $\gamma_e$ is still approximately given by the solution in the absence of magnetic fields, Equations~\eqref{eq:gammae-nomag} and \eqref{eq:xr-threshold}. However,~the density and the pressure are now given by Equations~\eqref{eq:rho-mag-weak} and \eqref{eq:pressure-mag-weak}, respectively. As~shown in  \mbox{Section~\ref{subsec:weakly-quantizing}}, these expansions in the weakly quantizing limit $\nu_{\rm max}\gg 1$ (including
the absence of magnetic field as a limiting case) actually remain very precise for $\nu_{\rm max}\sim 1$ and even at the onset of the strongly quantizing regime $\nu_{\rm max}=0$. Combining the solutions thus obtained with those presented in Section~\ref{sec:initial-composition-strongB}, the~full range of possible initial magnetic field strengths can  
be treated~analytically. 

\section{Magnetic Field Decay and Electron~Captures} 
\label{sec:heating}

\subsection{Onset of Electron~Captures}

The initial magnetic field decays on a very long time scale, say of the typical order of millions of years~\cite{pons2019}. The~compression of the crust thus occurs very slowly. When the pressure
of a matter element reaches some value $P_\beta(A,Z,B_\star)$, the~capture of an electron by nuclei $(A,Z)$ (in their ground state) opens.  The~daughter nuclei may be in an excited~state. 

The onset of electron captures by nuclei $(A,Z)$ is formally determined by the same condition 
irrespective of the magnetic field strength by requiring the constancy of the Gibbs free energy per nucleon at fixed temperature and pressure~\cite{fantina2015}. The~threshold electron Fermi energy is found to the first order in the fine-structure constant $\alpha$ from the condition:

\begin{equation}\label{eq:e-capture-gibbs-approx}
	\gamma_e + C_M \left(\frac{4\pi n_e}{3}\right)^{1/3} \alpha \lambda_e F(Z) = \gamma_e^{\beta} \, ,
\end{equation}
\begin{equation}\label{eq:def-F}
	F(Z)\equiv Z^{5/3}-(Z-1)^{5/3} + \frac{1}{3} Z^{2/3}\, ,
\end{equation}
\begin{equation}\label{eq:muebeta}
	\gamma_e^{\beta}\equiv -\frac{Q_{\rm EC}(A,Z,B_\star)}{m_e c^2} + 1 \, ,
\end{equation}
where 
we have introduced the $Q$-value (in vacuum) associated with electron capture by nuclei ($A,Z$):
\begin{equation}
	Q_{\rm EC}(A,Z,B_\star) = M^\prime(A,Z,B_\star)c^2-M^\prime(A,Z-1,B_\star)c^2-E_{\rm ex}(A,Z-1,B_\star)\, .
\end{equation}

\textls[20]{These $Q$-values can be obtained from the tabulated $Q$-values of $\beta$ decay by the following relation:}
\begin{equation}
	Q_{\rm EC}(A,Z,B_\star) = -Q_\beta(A,Z-1,B_\star)-E_{\rm ex}(A,Z-1,B_\star)\, .
\end{equation}

Here, $E_{\rm ex}(A,Z-1,B_\star)$ denotes the excitation energy of the daughter nucleus. Transitions to the ground state can be considered by setting $E_{\rm ex}(A,Z-1,B_\star)=0$.

\subsection{No Magnetic~Field}

\textls[-20]{In the absence of magnetic fields, the~threshold condition~(\ref{eq:e-capture-gibbs-approx}) can be solved exactly~\cite{chamelfantina2016}:}
\begin{equation}\label{eq:gammae-noB}
	\gamma_e = \sqrt{1+(x_r^\beta)^2}\, , 
\end{equation}
\begin{eqnarray}\label{eq:exact-xrbeta}
	x^\beta_r=\gamma_e^{\beta} \Biggl\{\sqrt{1-\biggl[1-\tilde F(Z)^2\biggr]/(\gamma_e^{\beta})^{2}}
	-\tilde F(Z)\Biggr\} \Biggl[1-\tilde F (Z)^2\Biggr]^{-1}\, ,
\end{eqnarray}
\begin{equation}\label{eq:Ftilde}
	\tilde F(Z)\equiv C_M \left(\frac{4}{9\pi}\right)^{1/3}\alpha F(Z)\, .
\end{equation}

The pressure $P_{\beta}(A,Z)$ and the density $\bar n_\beta(A,Z)$ at the onset of electron captures are then  
given by Equations~\eqref{eq:pressure-nomag} and \eqref{eq:rho-nomag}, respectively.

\subsection{Intermediate Magnetic~Field}

For $\nu_{\rm max}>0$, the~density and the pressure are obtained by substituting the solution~\eqref{eq:gammae-noB} for the threshold electron Fermi energy in the absence of magnetic field in Equations~\eqref{eq:rho-mag-weak} and \eqref{eq:pressure-mag-weak}, respectively.

\subsection{Strongly Quantizing Magnetic~Field}

In the strongly quantizing regime ($\nu_\text{max}=0$), Equation~\eqref{eq:e-capture-gibbs-approx} can be solved exactly from the general analytical solutions 
given in Section~\ref{sec:initial-composition-strongB}. Introducing
\begin{equation}
	\bar F(Z,B_\star)\equiv \frac{1}{3} C_M \alpha F(Z)\left(\frac{2 B_\star}{3\pi}\right)^{1/3} <0\, ,
\end{equation}
\begin{equation}
	\upsilon\equiv \frac{\gamma_e^{\beta}}{2 |\bar F(Z, B_\star)|^{3/2}}\, ,
\end{equation}
remarking that $\gamma_e^\beta>1$, the~solutions are given by the following formulas:

\begin{equation}\label{eq:exact-gammae-strongB}
	\gamma_e=\begin{cases}
		8|\bar F(Z, B_\star)|^{3/2}\, {\rm cosh}^3\left(\frac{1}{3}{\rm arccosh\,} \upsilon\right) & \text{if} \ \upsilon\geq 1\, ,\\
		8|\bar F(Z, B_\star)|^{3/2}\, \cos^3\left( \frac{1}{3}\arccos \upsilon\right) & \text{if} \ 0\leq \upsilon< 1\, . 
	\end{cases}
\end{equation}

The threshold pressure and density are respectively given by Equations~\eqref{eq:pressure-mag-strong} and \eqref{eq:rho-mag-strong}.

Let us recall that this solution is only valid under the assumption $\nu_{\rm max}=0$, which translates into a lower bound for the magnetic field $B_\star \geq B_{\star}^\beta$. To~find $B_{\star}^\beta$, 
we substitute Equations~\eqref{eq:gammae-mag-strong-cond} and \eqref{eq:ne-mag-strong-cond} in  Equation~\eqref{eq:e-capture-gibbs-approx}. This leads to
\begin{equation}\label{eq:Bstarbeta}
	B_{\star}^\beta=\frac{(\gamma_e^\beta)^2}{2}\biggl[1+\frac{C_M\alpha}{(3\pi)^{1/3}} F(Z)
	\biggr]^{-2}\, .
\end{equation}
We have assumed $B^\beta_{\star}\gg 1$ so that $\gamma_e \approx \sqrt{2 B_\star^\beta}$.

\subsection{Heat~Released}

The first electron capture does not release any significant heat since it essentially proceeds in quasiequilibrium. However, the~daughter nuclei (possibly in some excited state) are generally unstable and capture a second electron off-equilibrium thus depositing some heat at the same pressure $P_\beta(A,Z,B_\star)$. Ignoring the fraction of energy carried away by neutrinos, the~maximum amount of heat per nucleus $(A,Z)$ is given by
\begin{equation}
	\label{eq:heat-ocrust}
	\mathcal{Q}(A,Z,B_\star) = \mu(A,Z,B_\star)-\mu(A,Z-2,B_\star)\, .
\end{equation}

It is to be understood that the baryon chemical potentials must be evaluated at the same pressure. Expressing the electron Fermi energy associated with nuclei $(A,Z-2)$ as $\gamma_e+\delta \gamma_e$ with $\gamma_e$ given by the solution of Equation~\eqref{eq:e-capture-gibbs-approx}, and~expanding the pressure to the first order in $\delta \gamma_e$ leads to
\begin{equation} \label{eq:dgammae}
	\delta \gamma_e \approx \frac{1}{3}C_M \alpha \lambda_e \left(\frac{4\pi n_e}{3}\right)^{1/3}\left[ Z^{2/3}-(Z-2)^{2/3}\right]\, . 
\end{equation}

We have made use of the Gibbs--Duhem relation $dP_e=n_e m_e c^2 d\gamma_e$ and we have neglected terms of order $\alpha \delta \gamma_e$. Substituting Equation~\eqref{eq:dgammae} in Equation~\eqref{eq:heat-ocrust} and eliminating $\gamma_e$ using Equation~\eqref{eq:e-capture-gibbs-approx} lead to the following expression for the heat released per nucleus (keeping as before first-order terms):
\begin{eqnarray}
	\label{eq:heat-ocrust-approx}
	\mathcal{Q}(A,Z,B_\star) &&\approx  \mathcal{Q}^{(0)}(A,Z,B_\star) \nonumber \\
	&&-C_M \alpha m_e c^2 \lambda_e \left(\frac{4\pi A}{3 Z}\bar n \right)^{1/3}\biggl[ Z^{5/3}+(Z-2)^{5/3} -2(Z-1)^{5/3}\biggr] \, ,
\end{eqnarray} 
where the zeroth-order term is determined by nuclear data alone
\begin{eqnarray}
	\label{eq:heat-ocrust-approx-zeroth}
	\mathcal{Q}^{(0)}(A,Z,B_\star) &\equiv& 2 M^\prime(A,Z-1,B_\star)c^2+ 2 E_{\rm ex}(A,Z-1,B_\star) \nonumber \\&&-M^\prime(A,Z-2,B_\star)c^2-M^\prime(A,Z,B_\star)c^2 \, .
\end{eqnarray} 


Apart from the small electrostatic correction (the term proportional to the structure constant $C_M$), the~maximum heat released by electron captures is thus independent of whether the crust is solid or liquid. Unless~$B\gtrsim 10^{17}$~G~\cite{arteaga2011}, the~structure of nuclei remains essentially unchanged in the presence of a magnetic field so that $\mathcal{Q}(A,Z,B_\star) \approx \mathcal{Q}(A,Z)$. 

To estimate the heat in Equation~(\ref{eq:heat-ocrust}), we implicitly assumed  $\mu(A,Z-2,P_\beta,B_\star) < \mu(A,Z,P_\beta,B_\star)$, which generally holds for 
even $A$ nuclei, but~not necessarily for odd $A$ nuclei. In~the latter case, we typically have $Q_\beta(A,Z-1,B_\star) <Q_\beta(A,Z-2,B_\star)$. 
Using Equation~(\ref{eq:muebeta}), this implies that $\gamma_e^\beta(A,Z)<\gamma_e^\beta(A,Z-1)$. In~other words, as~the pressure reaches 
$P_\beta(A,Z,B_\star)$, the~nucleus ($A,Z$) decays, but~the daughter nucleus ($A,Z-1$) is actually stable against electron capture, and~therefore, 
no heat is released $\mathcal{Q}(A,Z,B_\star)=0$. The~daughter nucleus sinks deeper in the crust and only captures a second electron in quasi-equilibrium 
at pressure $P_\beta(A,Z-1,B_\star)>P_\beta(A,Z,B_\star)$.

\subsection{Neutron Delayed~Emission}

As discussed in~\cite{chamel2015b,fantina2016a}, the~first electron capture by the nucleus $(A,Z)$ may be accompanied by the emission of $\Delta N>0$ neutrons. 
The corresponding pressure $P_{\beta n}$ and baryon density $ n_{\beta n}$ are obtained from similar expressions as for electron captures except that the threshold electron Fermi energy $\gamma_e^\beta$ is now replaced by
\begin{equation}\label{eq:muedrip-acc}
	\gamma_e^{\beta n}= \frac{M^\prime(A-\Delta N,Z-1)c^2+E_{\rm ex}(A-\Delta N,Z-1)-M^\prime(A,Z)c^2+\Delta N m_n c^2}{m_e c^2} + 1 \, .
\end{equation}

Neutron emission will thus occur whenever $\gamma_e^{\beta n}(A,Z) < \gamma_e^\beta(A,Z)$. 

\section{Results and~Discussions}
\label{sec:results}

\subsection{Initial Composition of the Outer~Crust} 

The initial composition of the outer crust of a magnetar was determined in~\cite{mutafchieva2019} but only for a few selected magnetic field strengths, namely $B_\star=1000$, 2000, and~
3000. We have extended the calculations to the whole range of magnetic field strengths ranging from $B=0$ to $B=10^{17}$~G. To~this end, we have used the experimental atomic 
masses from the 2016 Atomic Mass Evaluation~\cite{ame2016} supplemented with the same microscopic atomic mass table HFB-24~\cite{goriely2013} from the BRUSLIB database\footnote{\url{http://www.astro.ulb.ac.be/bruslib/}, accessed on 9 June 2022}~\cite{bruslib}. 
The functional BSk24 underlying the model HFB-24 was also adopted to calculate the equation of state 
of the inner crust of a magnetar~\cite{mutafchieva2019}. This same functional was also applied to construct a unified equation of state for unmagnetized neutron  stars~\cite{pearson2018,pearson2020,pearson2022}, and~to calculate superfluid properties~\cite{ChamelAllard2021}. Results are publicly available on CompOSE\footnote{\url{https://compose.obspm.fr}, accessed on 9 June 2022}. This 
equation of state is consistent with the constraints 
inferred from analyses of the gravitational-wave signal from the binary neutron-star merger GW170817 and of its electromagnetic counterpart~\cite{perot2019}. As shown in~\cite{mutafchieva2019}, the~magnetic field has a negligible impact on the equation of the state of 
the inner crust and core of magnetars unless it exceeds about $10^{17}$~G. 

Depending on the strength of the magnetic field when the magnetar was born, different nuclides are expected to be produced in the outer crust. Changes in the composition compared 
to that obtained in~\cite{pearson2018} in the absence of the magnetic field are summarized in Tables~\ref{tab:trans-mag} and \ref{tab:composition}. The~mean baryon number densities 
and the pressure at the interface between adjacent crustal layers can be calculated for any magnetic field using the analytical formulas given in Section~\ref{sec:initial-composition}. 
The only nuclear inputs are embedded in the parameter $\gamma_e^{1\rightarrow 2}$ defined by Equation~\eqref{eq:mue12}. Values for this parameter are indicated in Table~\ref{tab:crust-layers}
for all possible~transitions. 

\begin{table}[H]
	\caption{Magnetic field strength $B_\star=B/B_{\rm rel}$ for the appearance (+) or the disappearance ($-$) of a nuclide in the outer crust of a~magnetar. }
	\label{tab:trans-mag}
	
	\begin{tabular}{m{6.5cm}<{\centering}m{6.5cm}<{\centering}}
		
		\toprule
		\textbf{Nuclide} & $\pmb{B_\star}$ \\ \midrule
		$^{66}$Ni($-$)& 67 \\  
		$^{88}$Sr(+)& 858 \\
		$^{126}$Ru(+)& 1023 \\ 
		$^{80}$Ni($-$)& 1072 \\ 
		$^{128}$Pd(+)& 1249 \\ 
		$^{78}$Ni($-$)& 1416  \\ 
		$^{64}$Ni($-$)&1669\\
		\bottomrule
	\end{tabular}
	
\end{table}

\begin{table}[H]\ContinuedFloat
	\caption{\textit{Cont}.}
	\label{tab:trans-mag}
	
	\begin{tabular}{m{6.5cm}<{\centering}m{6.5cm}<{\centering}}
		
		\toprule
		\textbf{Nuclide} & $\pmb{B_\star}$ \\ \midrule

		$^{124}$Zr(+) & 1872 \\ 
		$^{121}$Y($-$) & 1907 \\
		$^{132}$Sn(+)&1986\\ 
		$^{80}$Ni($-$)&2087 \\

		\bottomrule
	\end{tabular}
\end{table}

\vspace{-9pt}

\begin{table}[H] 
	\caption{Sequence of equilibrium nuclides with increasing depth (from top to bottom) in the outer crust of a magnetar for different magnetic field strengths. The~first row indicates values of $B_\star=B/B_{\rm rel}$ associated with a change of composition. Results are valid up to $B_ \star=2087$. 
	}
	\label{tab:composition}
	
	\begin{adjustwidth}{-\extralength}{0cm}
		\begin{tabular}{m{1.25cm}<{\centering}m{1.25cm}<{\centering}m{1.25cm}<{\centering}m{1.25cm}<{\centering}m{1.25cm}<{\centering}m{1.25cm}<{\centering}m{1.25cm}<{\centering}m{1.25cm}<{\centering}m{1.25cm}<{\centering}m{1.25cm}<{\centering}m{1.3cm}<{\centering}}
			\toprule
			& \textbf{67}                 & \textbf{858}                & \textbf{1023}               & \textbf{1072}              & \textbf{1249}              & \textbf{1416}              & \textbf{1669}                 & \textbf{1872}                & \textbf{1907}                 & \textbf{1986}                                             \\ \midrule
			$^{56}$Fe & $^{56}$Fe & $^{56}$Fe & $^{56}$Fe& $^{56}$Fe& $^{56}$Fe & $^{56}$Fe & $^{56}$Fe   & $^{56}$Fe  & $^{56}$Fe   & $^{56}$Fe                              \\ 
			$^{62}$Ni & $^{62}$Ni & $^{62}$Ni & $^{62}$Ni& $^{62}$Ni& $^{62}$Ni& $^{62}$Ni & $^{62}$Ni   & $^{62}${Ni}  & $^{62}$Ni   & $^{62}$Ni                               \\ 
			$^{64}${Ni} & $^{64}${Ni} & $^{64}${Ni} & $^{64}${Ni}& $^{64}${Ni}& $^{64}${Ni}& $^{64}${Ni} & --                 & --                & --                 & --                                             \\ 
			$^{66}${Ni} & --               & --               & --              & --              & --             & --              & --                & --               & --                & --                                            \\ 
			--               & --              & $^{88}${Sr} & $^{88}${Sr}& $^{88}${Sr}& $^{88}${Sr}& $^{88}${Sr} & $^{88}${Sr}   & $^{88}${Sr}  & $^{88}${Sr}   & $^{88}$Sr                               \\ 
			$^{86}${Kr} & $^{86}${Kr} & $^{86}${Kr} & $^{86}${Kr}& $^{86}${Kr}& $^{86}${Kr}& $^{86}${Kr} & $^{86}${Kr}   & $^{86}${Kr}  & $^{86}${Kr}   & $^{86}$Kr                               \\ 
			$^{84}${Se} & $^{84}${Se} & $^{84}${Se} & $^{84}${Se}& $^{84}${Se}& $^{84}${Se}& $^{84}${Se} & $^{84}${Se}   & $^{84}${Se}  & $^{84}${Se}   & $^{84}$Se                               \\ 
			$^{82}${Ge} & $^{82}${Ge} & $^{82}${Ge} & $^{82}${Ge}& $^{82}${Ge}& $^{82}${Ge}& $^{82}${Ge} & $^{82}${Ge}   & $^{82}${Ge}  & $^{82}${Ge}   & $^{82}$Ge                               \\ 
			--                & --              & --              & --             & --             & --             & --              & --                & --               & --                & $^{132}$Sn                              \\
			$^{80}${Zn} & $^{80}${Zn} & $^{80}${Zn} & $^{80}${Zn}& $^{80}${Zn}& $^{80}${Zn}& $^{80}${Zn} & $^{80}${Zn}   & $^{80}${Zn}  & $^{80}${Zn}   & $^{80}$Zn                               \\ 
			$^{78}${Ni} & $^{78}${Ni} & $^{78}${Ni} & $^{78}${Ni}& $^{78}${Ni}& $^{78}${Ni}& --              & --                & --               & --                & --                                            \\ 
			$^{80}${Ni} & $^{80}${Ni} & $^{80}${Ni} & $^{80}${Ni}& --             & --             & --              & --                & --               & --                & --                                            \\ 
			--                & --              & --              & --             & --             & $^{128}$Pd        & $^{128}$Pd         & $^{128}$Pd           & $^{128}$Pd          & $^{128}$Pd           & $^{128}$Pd                                       \\ 
			--                & --              & --              & $^{126}$Ru        & $^{126}$Ru        & $^{126}$Ru        &  $^{126}$Ru        & $^{126}$Ru           & $^{126}$Ru          & $^{126}$Ru           & $^{126}$Ru                                       \\ 
			$^{124}$Mo         & $^{124}$Mo         & $^{124}$Mo         & $^{124}$Mo        & $^{124}$Mo        & $^{124}$Mo        & $^{124}$Mo         & $^{124}$Mo           & $^{124}$Mo          & $^{124}$Mo           & $^{124}$Mo                                       \\ 
			$^{122}$Zr         & $^{122}$Zr         & $^{122}$Zr         & $^{122}$Zr        & $^{122}$Zr        & $^{122}$Zr        & $^{122}$Zr         & $^{122}$Zr           & $^{122}$Zr          & $^{122}$Zr           & $^{122}$Zr                                       \\ 
			--                & --              & --              & --             & --             & --             & --              & --                & $^{124}$Zr          & $^{124}$Zr           & $^{124}$Zr                                       \\ 
			$^{121}$Y          & $^{121}$Y          & $^{121}$Y          & $^{121}$Y         & $^{121}$Y         & $^{121}$Y         & $^{121}$Y          & $^{121}$Y            & $^{121}$Y           & --                & --                                            \\ 
			$^{120}$Sr         & $^{120}$Sr         & $^{120}$Sr         & $^{120}$Sr        & $^{120}$Sr        & $^{120}$Sr        & $^{120}$Sr         & $^{120}$Sr           & $^{120}$Sr          & $^{120}$Sr           & $^{120}$Sr                                       \\ 
			$^{122}$Sr         & $^{122}$Sr         & $^{122}$Sr         & $^{122}$Sr        & $^{122}$Sr        & $^{122}$Sr        & $^{122}$Sr         & $^{122}$Sr           & $^{122}$Sr          & $^{122}$Sr           & $^{122}$Sr                                       \\ 
			$^{124}$Sr         & $^{124}$Sr         & $^{124}$Sr         & $^{124}$Sr        & $^{124}$Sr        & $^{124}$Sr        & $^{124}$Sr         & $^{124}$Sr           & $^{124}$Sr          & $^{124}$Sr           & $^{124}$Sr                                       \\
			\bottomrule                      
		\end{tabular}      
	\end{adjustwidth}
\end{table}

\vspace{-9pt}
\begin{table}[H] 
	\caption{Values of the nuclear parameter $\gamma_e^{1\rightarrow 2}$ from which the pressure and the densities at the boundary between adjacent layers in the outer crust of a magnetar can be~calculated. }
	\label{tab:crust-layers}
	
	\begin{tabular}{m{6.5cm}<{\centering}m{6.5cm}<{\centering}}
		
		\toprule
		$\pmb{\gamma_e^{1\rightarrow 2}}$      & \textbf{Interface} 
		\\  \midrule
		1.8908                           & $^{56}\mathrm{Fe}$ -- ${}^{62}\mathrm{Ni}$   \\
		4.8972                           & ${}^{62}\mathrm{Ni}$ -- ${}^{64}\mathrm{Ni}$   \\
		8.6863                           & ${}^{62}\mathrm{Ni}$ -- ${}^{88}\mathrm{Sr}$   \\
		8.1312                           & ${}^{64}\mathrm{Ni}$ -- ${}^{66}\mathrm{Ni}$   \\
		9.3317                           & ${}^{64}\mathrm{Ni}$ -- ${}^{86}\mathrm{Kr}$   \\
		18.098                           & ${}^{64}\mathrm{Ni}$ -- ${}^{88}\mathrm{Sr}$   \\
		12.155                           & ${}^{66}\mathrm{Ni}$ -- ${}^{86}\mathrm{Kr}$   \\
		10.044                           & ${}^{86}\mathrm{Kr}$ -- ${}^{84}\mathrm{Se}$   \\
		5.5622                           & ${}^{88}\mathrm{Sr}$ -- ${}^{86}\mathrm{Kr}$   \\
		15.330                           & ${}^{84}\mathrm{Se}$ -- ${}^{82}\mathrm{Ge}$   \\
		20.519                           & ${}^{82}\mathrm{Ge}$ -- ${}^{80}\mathrm{Zn}$   \\
		38.310                           & ${}^{82}\mathrm{Ge}$ -- ${}^{132}\mathrm{Sn}$   \\
		25.926                           & ${}^{80}\mathrm{Zn}$ -- ${}^{78}\mathrm{Ni}$   \\
		\bottomrule
	\end{tabular}
	
\end{table}

\vspace{-9pt}
\begin{table}[H]\ContinuedFloat
	\caption{\textit{Cont}.}
	\label{tab:crust-layers}
	
	\begin{tabular}{m{6.5cm}<{\centering}m{6.5cm}<{\centering}}
		
		\toprule
		$\pmb{\gamma_e^{1\rightarrow 2}}$      & \textbf{Interface }                                \\  \midrule
		37.083                           & ${}^{80}\mathrm{Zn}$ -- ${}^{128}\mathrm{Pd}$   \\
		$-33.289$                        & ${}^{132}\mathrm{Sn}$ -- ${}^{80}\mathrm{Zn}$   \\
		32.978                           & ${}^{78}\mathrm{Ni}$ -- ${}^{80}\mathrm{Ni}$   \\
		$-409.21$                        & ${}^{78}\mathrm{Ni}$ -- ${}^{128}\mathrm{Pd}$   \\
		48.055                           & ${}^{78}\mathrm{Ni}$ -- ${}^{126}\mathrm{Ru}$   \\
		45.122                           & ${}^{80}\mathrm{Ni}$ -- ${}^{124}\mathrm{Mo}$   \\
		218.54                           & ${}^{80}\mathrm{Ni}$ -- ${}^{126}\mathrm{Ru}$   \\
		30.039                           & ${}^{128}\mathrm{Pd}$ -- ${}^{126}\mathrm{Ru}$   \\
		32.010                           & ${}^{126}\mathrm{Ru}$ -- ${}^{124}\mathrm{Mo}$   \\
		37.441                           & ${}^{124}\mathrm{Mo}$ -- ${}^{122}\mathrm{Zr}$   \\
		40.352                           & ${}^{122}\mathrm{Zr}$ -- ${}^{121}\mathrm{Y}$   \\
		42.293                           & ${}^{122}\mathrm{Zr}$ -- ${}^{124}\mathrm{Zr}$   \\
		1.8541                           & ${}^{124}\mathrm{Zr}$ -- ${}^{121}\mathrm{Y}$   \\
		39.031                           & ${}^{124}\mathrm{Zr}$ -- ${}^{120}\mathrm{Sr}$   \\
		40.786                           & ${}^{121}\mathrm{Y}$ -- ${}^{120}\mathrm{Sr}$   \\
		44.857                           & ${}^{120}\mathrm{Sr}$ -- ${}^{122}\mathrm{Sr}$   \\
		47.747                           & ${}^{122}\mathrm{Sr}$ -- ${}^{124}\mathrm{Sr}$   \\
		
		\bottomrule
	\end{tabular}
\end{table}

\subsection{Heating}

We have estimated the heat released by electron captures and their location using the experimental atomic masses and the $Q_\beta$ values (including the recommended ones) from the 2016 Atomic Mass Evaluation~\cite{ame2016} 
supplemented with the atomic mass model HFB-24~\cite{goriely2013}. 
We have taken excitation energies from the Nuclear Data section of the International Atomic Energy Agency
website\footnote{\url{https://www-nds.iaea.org/relnsd/NdsEnsdf/QueryForm.html}, accessed on 9 June 2022} following the Gamow--Teller selection rules, namely that the parity of the final state is the same as that of the initial state, whereas the total angular momentum 
$J$ can either remain unchanged or vary by $\pm \hbar $  (excluding transitions from $J=0$ to $J=0$). 

The threshold density and pressure for the onset of each electron capture, as~well as the amount of heat deposited, can be calculated for 
any magnetic field strength 
from the analytical formulas presented in Section~\ref{sec:heating} using the parameters indicated in Table~\ref{tab:reac-gs} for ground-state to ground-state transitions, 
in Table~\ref{tab:reac-exc} for ground-state to excited state transitions, and~in Table~\ref{tab:reac-CO} for transitions involving light elements that could have been accreted
from the interstellar medium. Full numerical results are freely available in~\cite{zenodo22}.

\begin{table}[H] 
	\caption{Values of the nuclear parameters $\gamma_e^{\beta}$ and $\mathcal{Q}^{(0)}$ (considering ground-state to ground-state transitions) from which the threshold density and pressure for the onset of electron captures, as~well as the heat released can be calculated. The~symbol ($\star$) is used to distinguish reactions for which theoretical atomic masses were~needed. 
	}
	\label{tab:reac-gs}
	
	\begin{tabular}{m{3cm}<{\centering}m{6.6cm}<{\centering}m{3cm}<{\centering}}
		
		\toprule
		$\pmb{\gamma_e^{\beta}}$      & \textbf{Reaction} &     $\pmb{\mathcal{Q}^{(0)}}$  \textbf{(MeV)} \\ \midrule
		8.232 & $^{56}\mathrm{Fe} \rightarrow  {}^{56}\mathrm{Cr} -2e^-+2\nu_e $ &  2.069 \\
		18.867 & $^{56}\mathrm{Cr} \rightarrow {}^{56}\mathrm{Ti} -2e^-+2\nu_e  $ &  2.295 \\
		29.313 & $^{56}\mathrm{Ti} \rightarrow {}^{56}\mathrm{Ca} -2e^-+2\nu_e  $ &  3.514 \\
		43.710 & $^{56}\mathrm{Ca} \rightarrow {}^{56}\mathrm{Ar} -2e^-+2\nu_e \ (\star) $ &  2.045 \\
		11.415 & $^{62}\mathrm{Ni} \rightarrow {}^{62}\mathrm{Fe} -2e^-+2\nu_e $ &  2.776 \\
		21.262 & $^{62}\mathrm{Fe} \rightarrow {}^{62}\mathrm{Cr} -2e^-+2\nu_e $ &  2.725 \\
		31.174 & $^{62}\mathrm{Cr} \rightarrow {}^{62}\mathrm{Ti} -2e^-+2\nu_e $ &  2.442 \\
		41.470 & $^{62}\mathrm{Ti} \rightarrow {}^{62}\mathrm{Ca} -2e^-+2\nu_e \ (\star) $ &  2.490 \\
		\bottomrule
	\end{tabular}
	
\end{table}

\begin{table}[H]\ContinuedFloat
	\caption{\textit{Cont}.}
	\label{tab:reac-gs}
	
	\begin{tabular}{m{3cm}<{\centering}m{6.6cm}<{\centering}m{3cm}<{\centering}}
		
		\toprule
		$\pmb{\gamma_e^{\beta}}$      & \textbf{Reaction} &     $\pmb{\mathcal{Q}^{(0)}}$  \textbf{(MeV)} \\ \midrule
		15.299 & $^{64}\mathrm{Ni} \rightarrow {}^{64}\mathrm{Fe} -2e^-+2\nu_e $ &  2.484 \\
		24.445 & $^{64}\mathrm{Fe} \rightarrow {}^{64}\mathrm{Cr} -2e^-+2\nu_e $ &  2.471 \\
		34.581 & $^{64}\mathrm{Cr} \rightarrow {}^{64}\mathrm{Ti} -2e^-+2\nu_e $ &  1.865 \\
		19.782 & $^{66}\mathrm{Ni} \rightarrow {}^{66}\mathrm{Fe} -2e^-+2\nu_e $ &  3.257 \\
		27.062 & $^{66}\mathrm{Fe} \rightarrow {}^{66}\mathrm{Cr} -2e^-+2\nu_e $ &  1.287 \\
		38.397 & $^{66}\mathrm{Cr} \rightarrow {}^{66}\mathrm{Ti} -2e^-+2\nu_e \ (\star) $ &  3.540 \\
		15.938 & $^{86}\mathrm{Kr} \rightarrow {}^{86}\mathrm{Se} -2e^-+2\nu_e $ &  2.504 \\
		23.585 & $^{86}\mathrm{Se} \rightarrow {}^{86}\mathrm{Ge} -2e^-+2\nu_e $ &  1.979 \\
		30.980 & $^{86}\mathrm{Ge} \rightarrow {}^{86}\mathrm{Zn} -2e^-+2\nu_e \ (\star) $ &  2.310 \\
		38.926 & $^{86}\mathrm{Zn} \rightarrow {}^{86}\mathrm{Ni} -2e^-+2\nu_e \ (\star) $ &  2.320 \\
		20.754 & $^{84}\mathrm{Se} \rightarrow {}^{84}\mathrm{Ge} -2e^-+2\nu_e $ &  2.389 \\
		28.517 & $^{84}\mathrm{Ge} \rightarrow {}^{84}\mathrm{Zn} -2e^-+2\nu_e $ &  1.903 \\
		36.362 & $^{84}\mathrm{Zn} \rightarrow {}^{84}\mathrm{Ni} -2e^-+2\nu_e \ (\star) $ &  2.390 \\
		25.431 & $^{82}\mathrm{Ge} \rightarrow {}^{82}\mathrm{Zn} -2e^-+2\nu_e $ &  1.868 \\
		34.256 & $^{82}\mathrm{Zn} \rightarrow {}^{82}\mathrm{Ni} -2e^-+2\nu_e \ (\star) $ &  2.154 \\
		44.640 & $^{82}\mathrm{Ni} \rightarrow {}^{82}\mathrm{Fe} -2e^-+2\nu_e \ (\star) $ &  2.080 \\
		31.233 & $^{80}\mathrm{Zn} \rightarrow {}^{80}\mathrm{Ni} -2e^-+2\nu_e $ &  1.879 \\
		39.435 & $^{78}\mathrm{Ni} \rightarrow {}^{78}\mathrm{Fe} -2e^-+2\nu_e \ (\star) $ &  2.070 \\
		41.842 & $^{124}\mathrm{Mo} \rightarrow {}^{124}\mathrm{Zr} -2e^-+2\nu_e \ (\star) $ &  2.920 \\
		42.722 & $^{122}\mathrm{Zr} \rightarrow {}^{122}\mathrm{Sr} -2e^-+2\nu_e \ (\star) $ &  0.790 \\
		11.397 & $^{88}\mathrm{Sr} \rightarrow {}^{88}\mathrm{Kr} -2e^-+2\nu_e $ &  2.395 \\
		18.564 & $^{88}\mathrm{Kr} \rightarrow {}^{88}\mathrm{Se} -2e^-+2\nu_e $ &  2.144 \\
		26.761 & $^{88}\mathrm{Se} \rightarrow {}^{88}\mathrm{Ge} -2e^-+2\nu_e $ &  2.582 \\
		34.601 & $^{88}\mathrm{Ge} \rightarrow {}^{88}\mathrm{Zn} -2e^-+2\nu_e \ (\star) $ &  2.740 \\
		36.695 & $^{126}\mathrm{Ru} \rightarrow {}^{126}\mathrm{Mo} -2e^-+2\nu_e \ (\star) $ &  1.860 \\
		34.444 & $^{128}\mathrm{Pd} \rightarrow {}^{128}\mathrm{Ru} -2e^-+2\nu_e \ (\star) $ &  2.290 \\
		28.662 & $^{132}\mathrm{Sn} \rightarrow {}^{132}\mathrm{Cd} -2e^-+2\nu_e $ &  1.987 \\
		33.237 & $^{132}\mathrm{Cd} \rightarrow {}^{132}\mathrm{Pd} -2e^-+2\nu_e \ (\star) $ &  2.863 \\
		38.084 & $^{132}\mathrm{Pd} \rightarrow {}^{132}\mathrm{Ru} -2e^-+2\nu_e \ (\star) $ &  2.940 \\
		
		\bottomrule
	\end{tabular}
\end{table}

\vspace{-9pt}
\begin{table}[H] 
	\caption{Same as in Table~\ref{tab:reac-gs} but considering ground-state to excited state transitions (for which the excitation energy is experimentally known and $E_{\textrm{exc}} > 0$). 
		\label{tab:reac-exc}}
	
	\begin{tabular}{m{3cm}<{\centering}m{6.6cm}<{\centering}m{3cm}<{\centering}}
		\toprule
		$\pmb{\gamma_e^{\beta}}$      & \textbf{Reaction} &    $\pmb{\mathcal{Q}^{(0)}}$  \textbf{(MeV)} \\ \midrule
		8.448 & $^{56}\mathrm{Fe} \rightarrow {}^{56}\mathrm{Cr} -2e^-+2\nu_e $ &  2.290 \\
		12.405 & $^{62}\mathrm{Ni} \rightarrow {}^{62}\mathrm{Fe} -2e^-+2\nu_e $ &  3.788 \\
		20.726 & $^{86}\mathrm{Kr} \rightarrow {}^{86}\mathrm{Se} -2e^-+2\nu_e $ &  7.397 \\
		31.260 & $^{82}\mathrm{Ge} \rightarrow {}^{82}\mathrm{Zn} -2e^-+2\nu_e $ &  7.825 \\
		15.764 & $^{88}\mathrm{Sr} \rightarrow {}^{88}\mathrm{Kr} -2e^-+2\nu_e $ &  6.858 \\
		22.290 & $^{88}\mathrm{Kr} \rightarrow {}^{88}\mathrm{Se} -2e^-+2\nu_e $ &  5.951 \\
		31.010 & $^{132}\mathrm{Sn} \rightarrow {}^{132}\mathrm{Cd} -2e^-+2\nu_e $ &  4.387 \\
		
		\bottomrule
	\end{tabular}
	
\end{table}

\vspace{-9pt}
\begin{table}[H] 
	\caption{Same as in Tables~\ref{tab:reac-gs} but considering reactions involving carbon and oxygen. 
		\label{tab:reac-CO}}
	
	\begin{tabular}{m{3cm}<{\centering}m{6.6cm}<{\centering}m{3cm}<{\centering}}
		\toprule
		$\pmb{\gamma_e^{\beta}}$      & \textbf{Reaction} &    $\pmb{\mathcal{Q}^{(0)}}$  \textbf{(MeV)} \\ \midrule
		21.393 & $^{16}\mathrm{O} \rightarrow {}^{16}\mathrm{C} -2e^-+2\nu_e $ &  2.411 \\
		27.163 & $^{12}\mathrm{C} \rightarrow {}^{12}\mathrm{Be} -2e^-+2\nu_e $ &  1.661 \\
		
		\bottomrule
	\end{tabular}
	
\end{table}

\begin{figure}[H]
	
	\includegraphics[width=10.5cm]{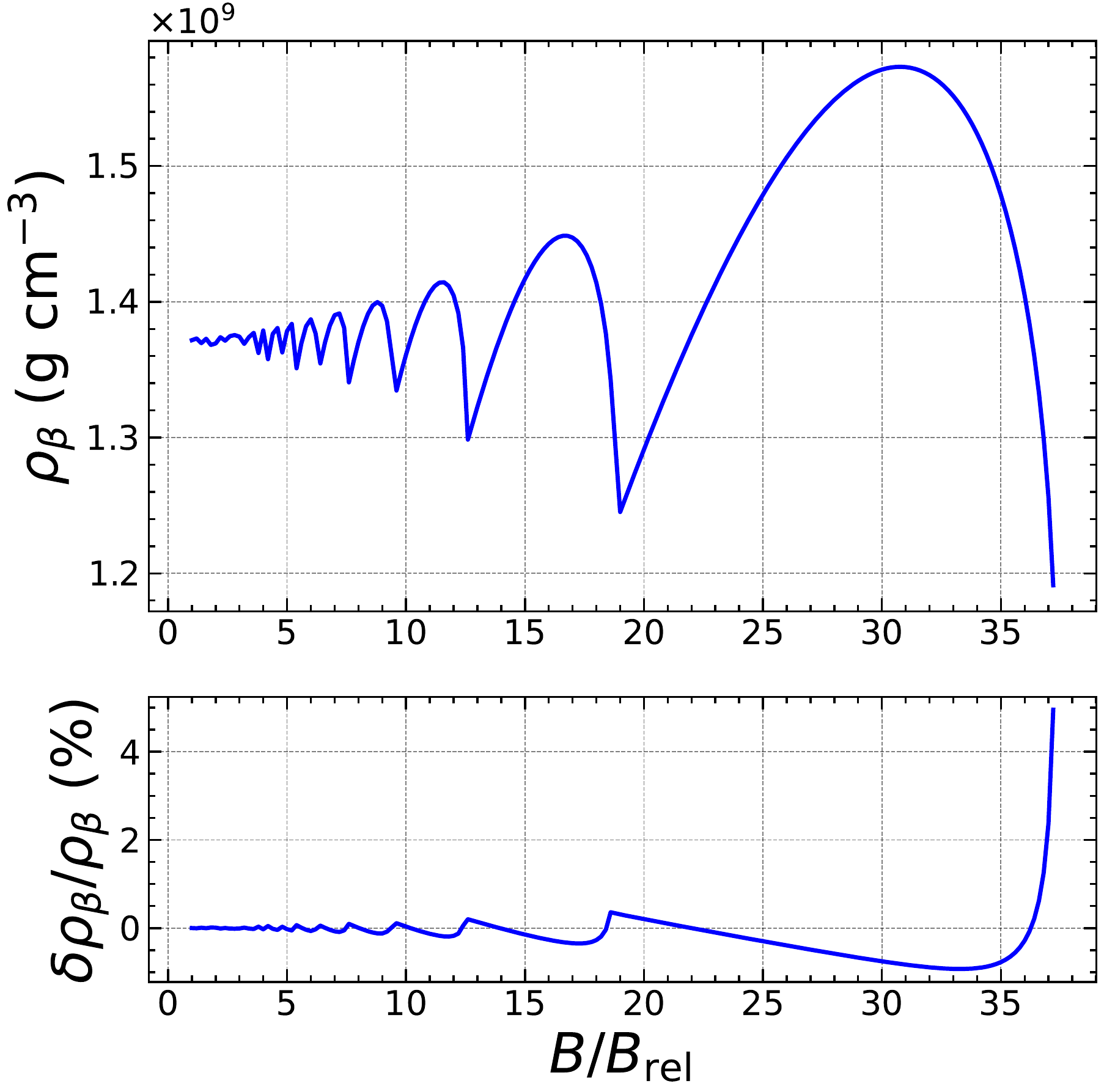}
	\caption{\textbf{Top panel}: Exact threshold density (in cgs units) for the onset of electron captures by $^{56}$Fe as a function of the magnetic field strength $B_\star=B/B_{\rm rel}$ up to the onset of the strongly quantizing regime. \textbf{Bottom panel}: Relative error (in \%) of the approximate analytical expression. 
	}
	\label{fig4}
\end{figure}

\begin{figure}[H]
	
	\includegraphics[width=10.5cm]{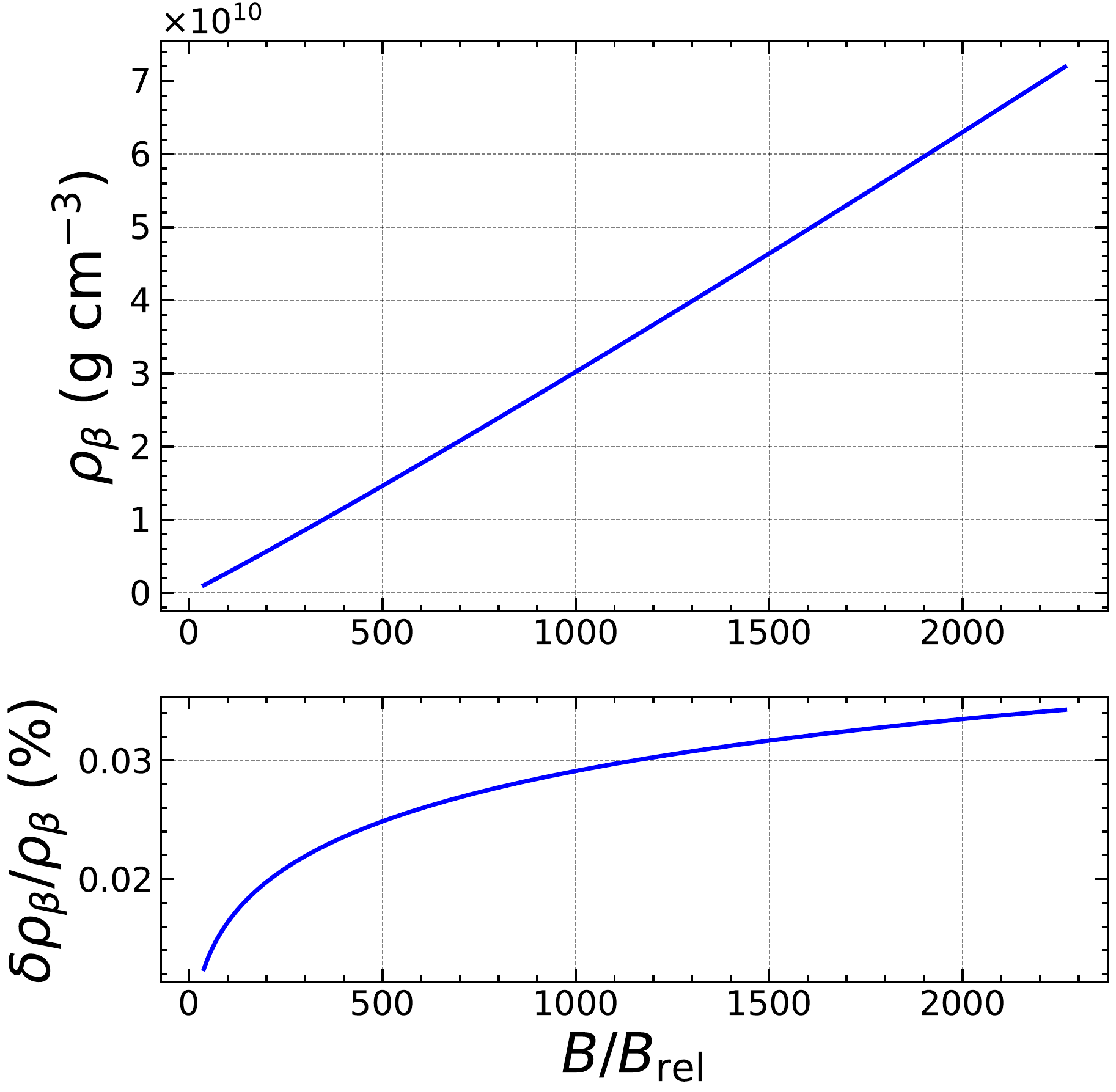}
	\caption{Same as Figure~\ref{fig4} in the strongly quantizing regime. 
	}
	\label{fig5}
\end{figure}

\begin{figure}[H]
	
	\includegraphics[width=10.5cm]{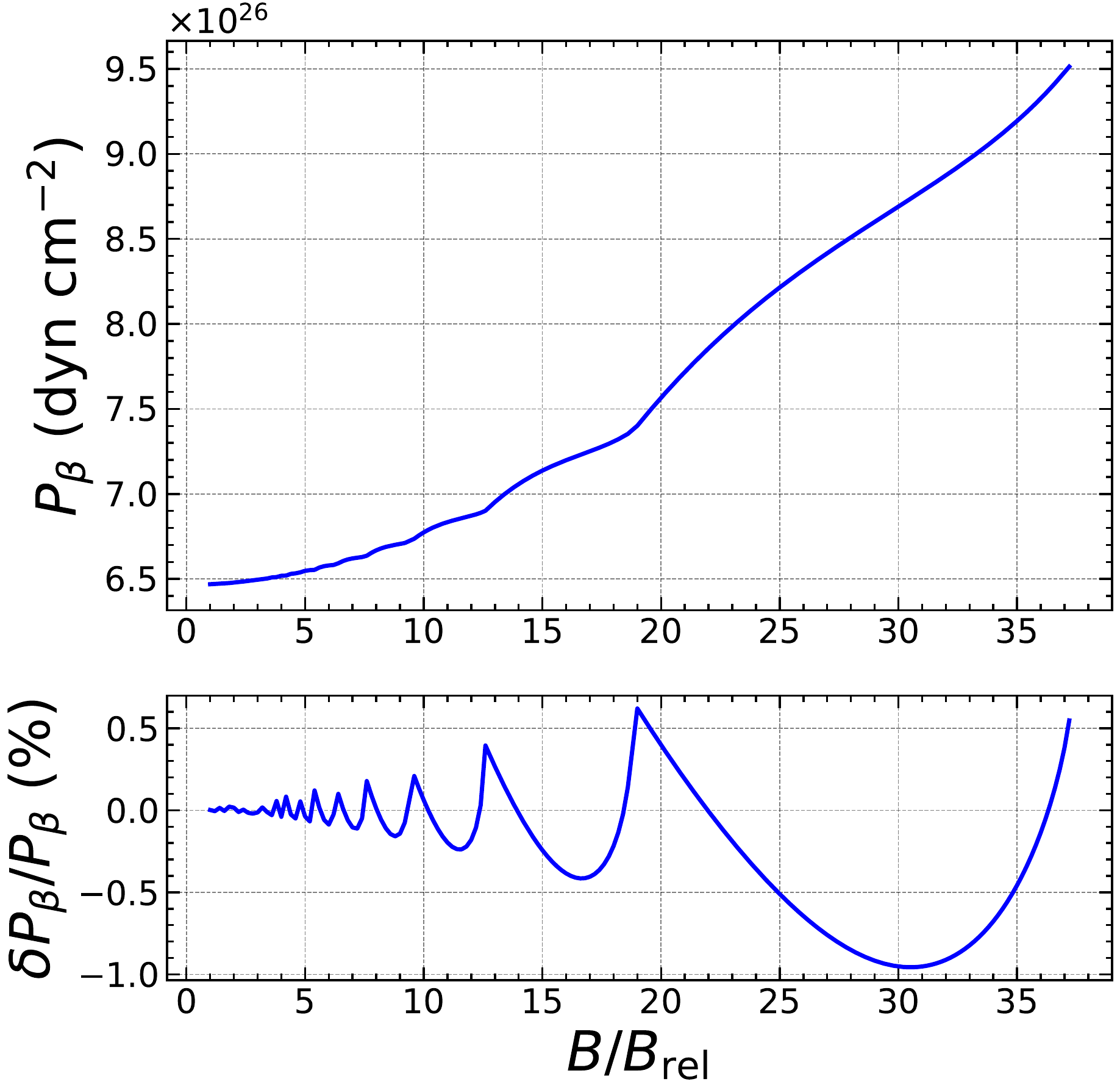}
	\caption{\textbf{Top panel}: Exact threshold pressure (in cgs units) for the onset of electron captures by $^{56}$Fe as a function of the magnetic field strength $B_\star=B/B_{\rm rel}$ up to the onset of the strongly quantizing regime. \textbf{Bottom panel}: Relative error (in \%) of the approximate analytical expression. 
	}
	\label{fig2}
\end{figure}

\begin{figure}[H]
	
	\includegraphics[width=10.5cm]{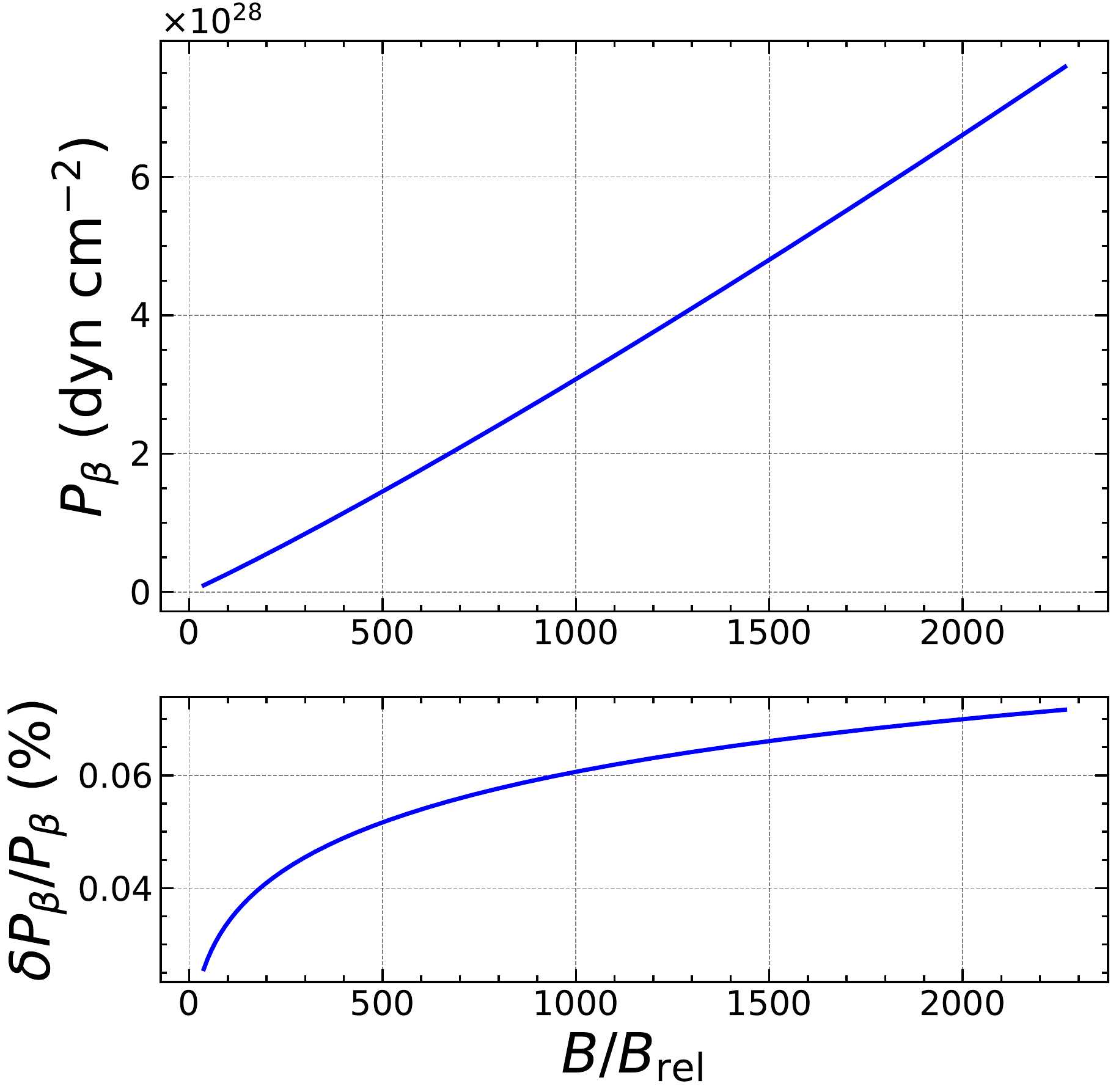}
	\caption{Same as Figure~\ref{fig2} in the strongly quantizing regime. 
	}
	\label{fig3}
\end{figure}

\begin{figure}[H]
	
	\includegraphics[width=10.5cm]{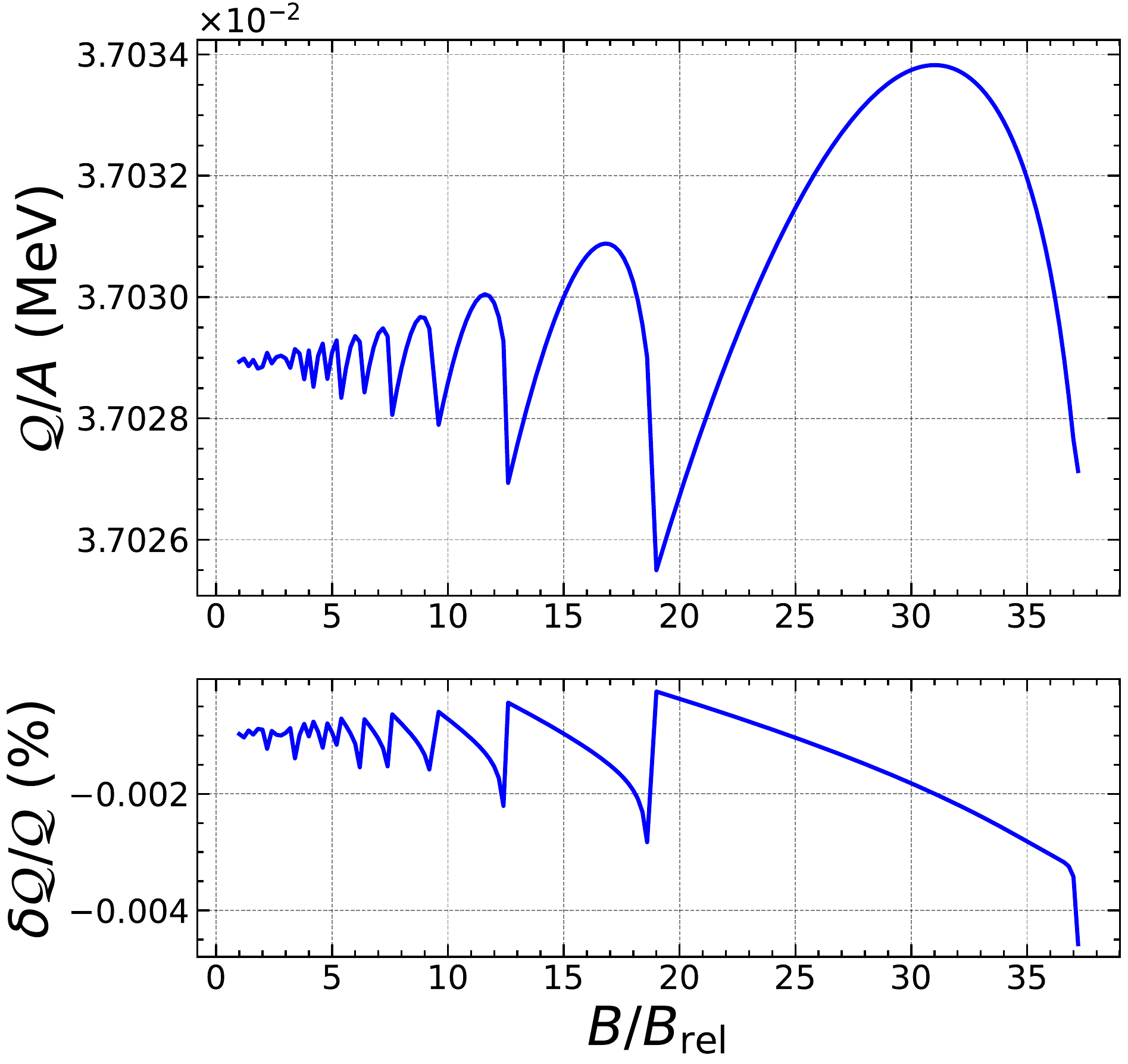}
	\caption{\textbf{Top panel}: Exact heat released per nucleon from electron captures by $^{56}$Fe as a function of the magnetic field strength $B_\star=B/B_{\rm rel}$ up to the onset of the strongly quantizing regime. \textbf{Bottom panel}: Relative error (in \%) of the approximate analytical expression. 
	}
	\label{fig6}
\end{figure}

\begin{figure}[H]
	
	\includegraphics[width=10.5cm]{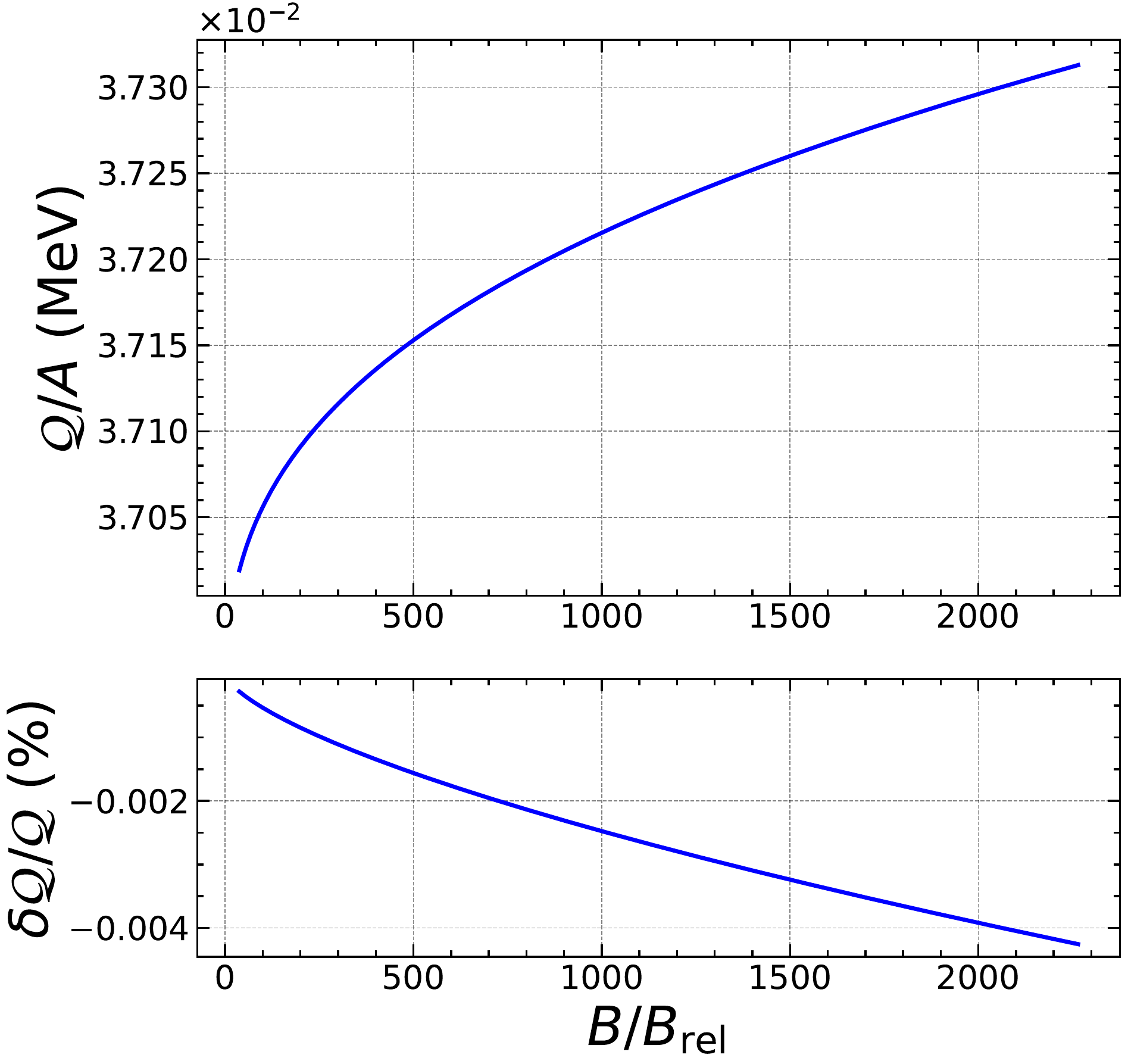}
	\caption{Same as Figure~\ref{fig6} in the strongly quantizing regime. 
	}
	\label{fig7}
\end{figure}

\begin{figure}[H]
	
	\includegraphics[width=10.5cm]{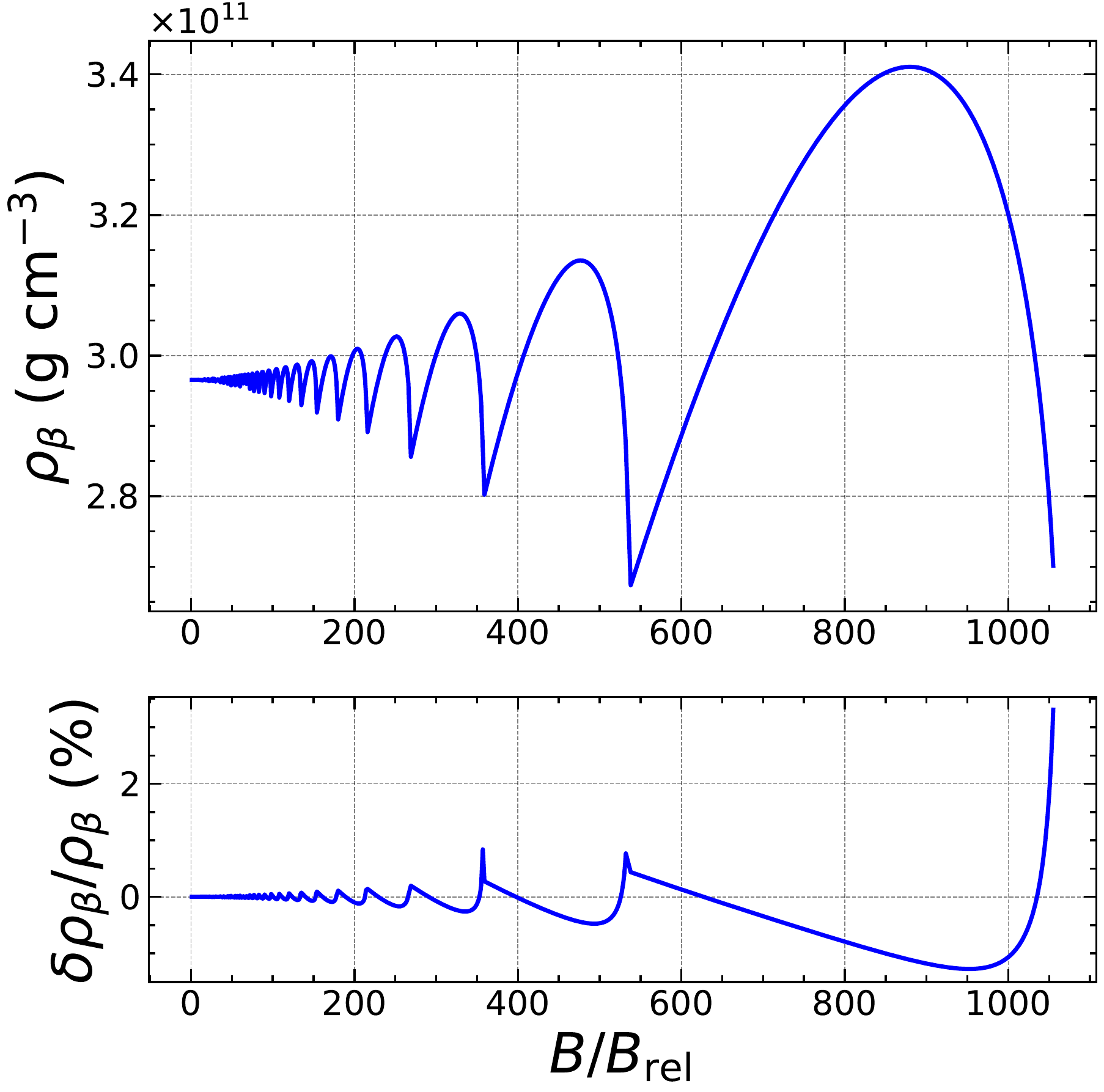}
	\caption{\textbf{Top panel}: Exact threshold density (in cgs units) for the onset of electron captures by $^{122}$Zr as a function of the magnetic field strength $B_\star=B/B_{\rm rel}$ up to the onset of the strongly quantizing regime. \textbf{Bottom panel}: Relative error (in \%) of the approximate analytical expression. 
	}
	\label{fig8}
\end{figure}

\begin{figure}[H]
	
	\includegraphics[width=10.5cm]{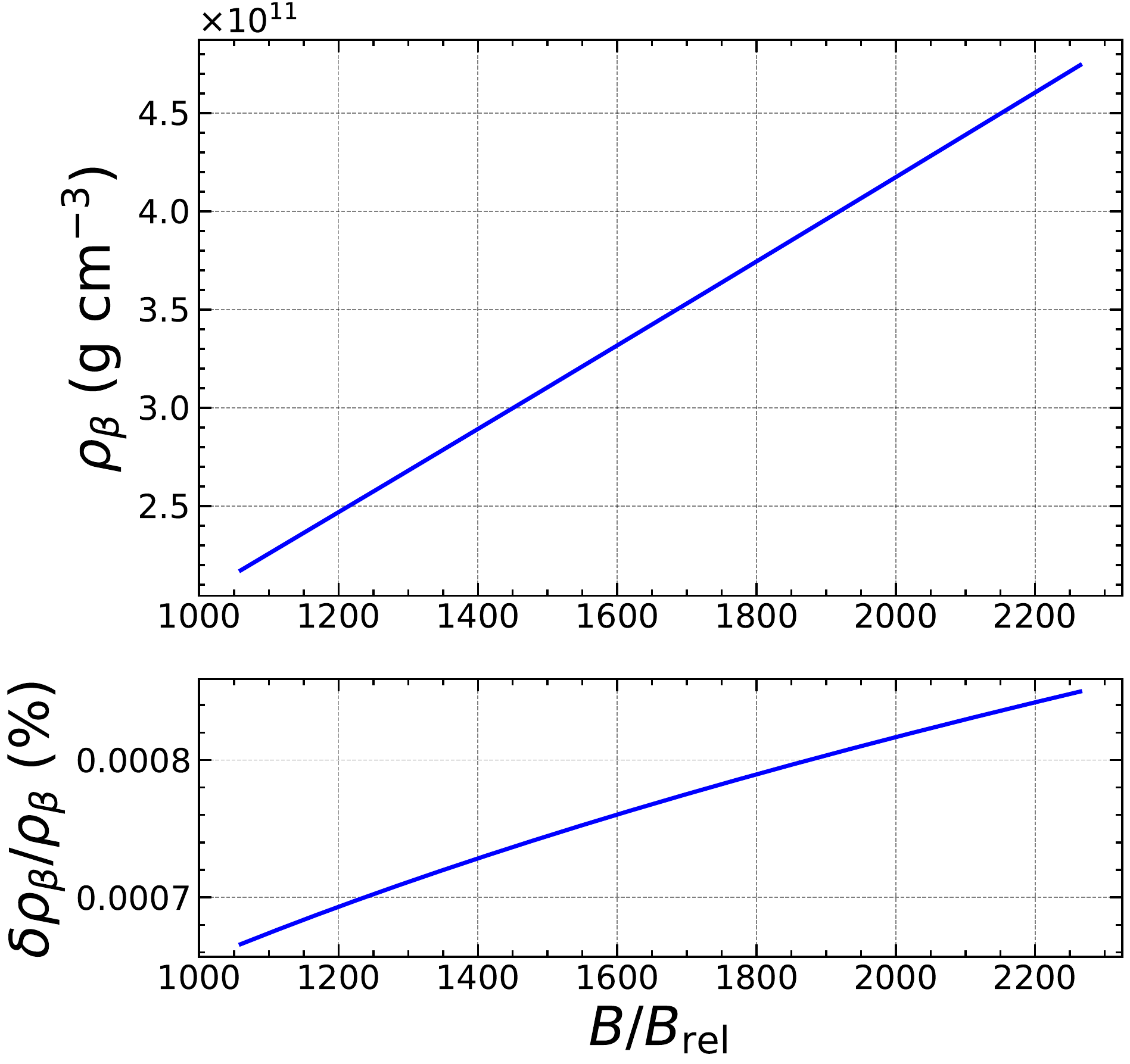}
	\caption{Same as Figure~\ref{fig8} in the strongly quantizing regime. 
	}
	\label{fig9}
\end{figure}

\begin{figure}[H]
	
	\includegraphics[width=10.5cm]{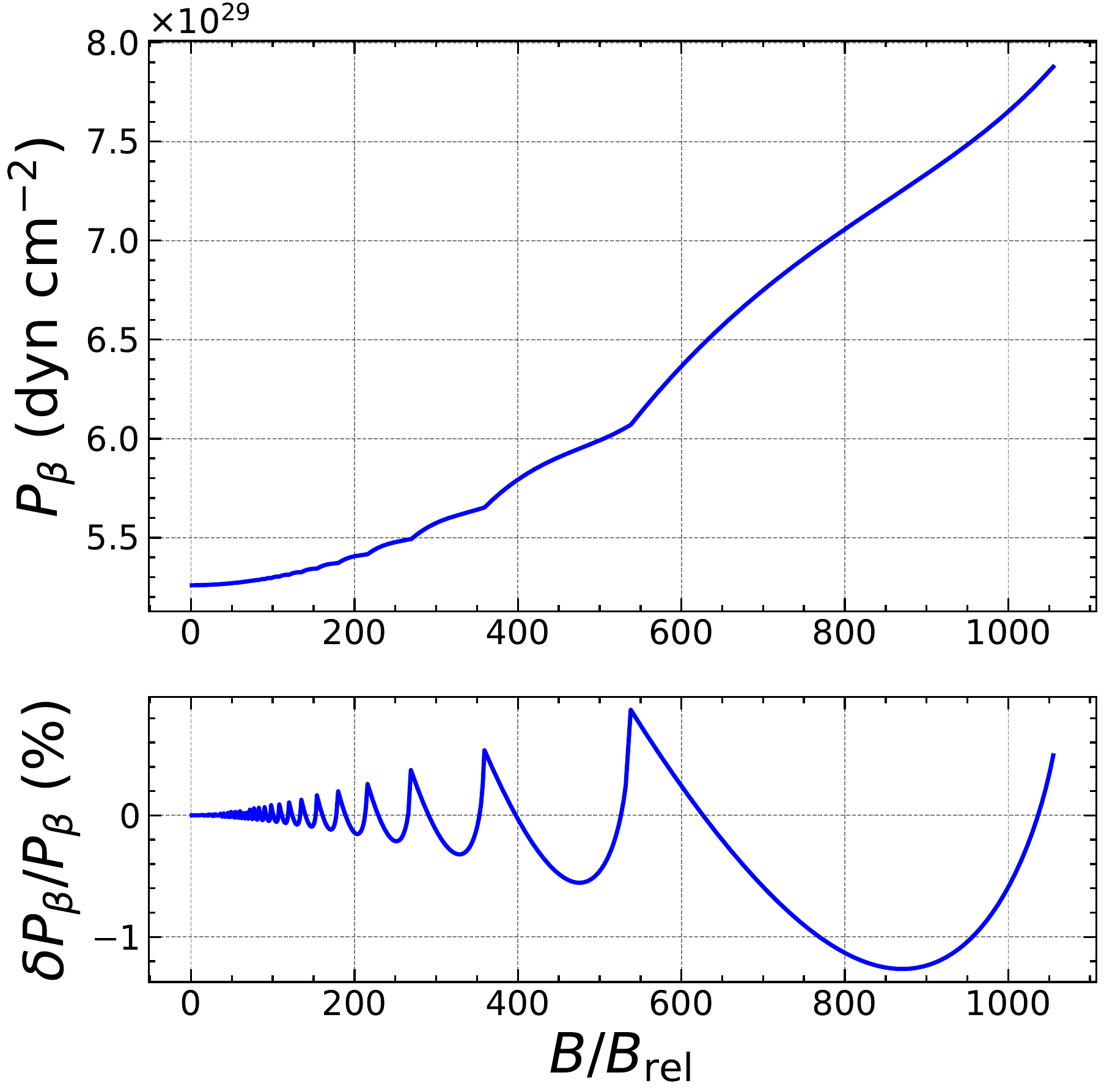}
	\caption{\textbf{Top panel}: Exact threshold pressure (in cgs units) for the onset of electron captures by $^{122}$Zr as a function of the magnetic field strength $B_\star=B/B_{\rm rel}$ up to the onset of the strongly quantizing regime. \textbf{Bottom panel}: Relative error (in \%) of the approximate analytical expression. 
	}
	\label{fig10}
\end{figure}

\begin{figure}[H]
	
	\includegraphics[width=10.5cm]{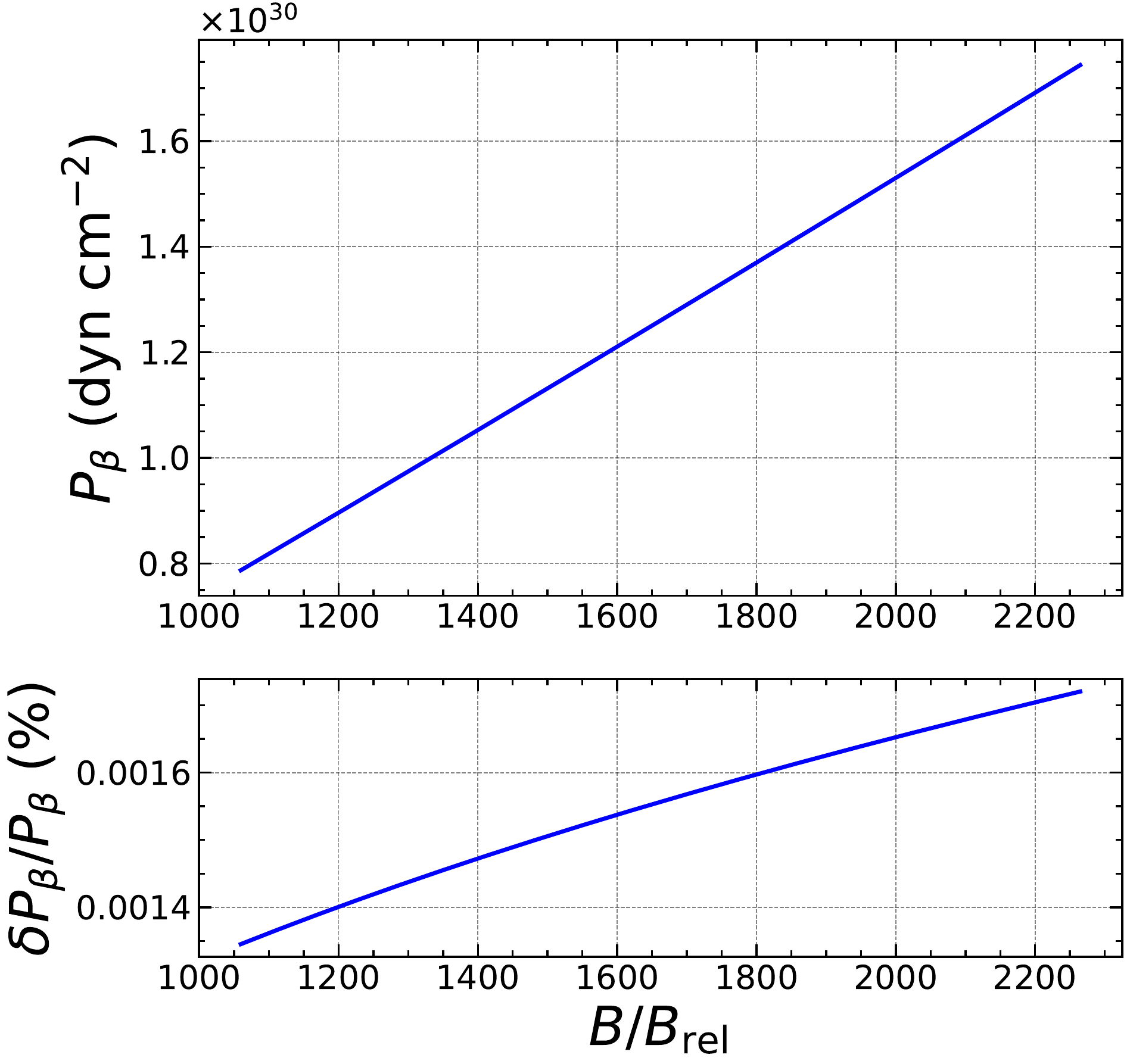}
	\caption{Same as Figure~\ref{fig10} in the strongly quantizing regime. 
	}
	\label{fig11}
\end{figure}

\begin{figure}[H]
	
	\includegraphics[width=10.5cm]{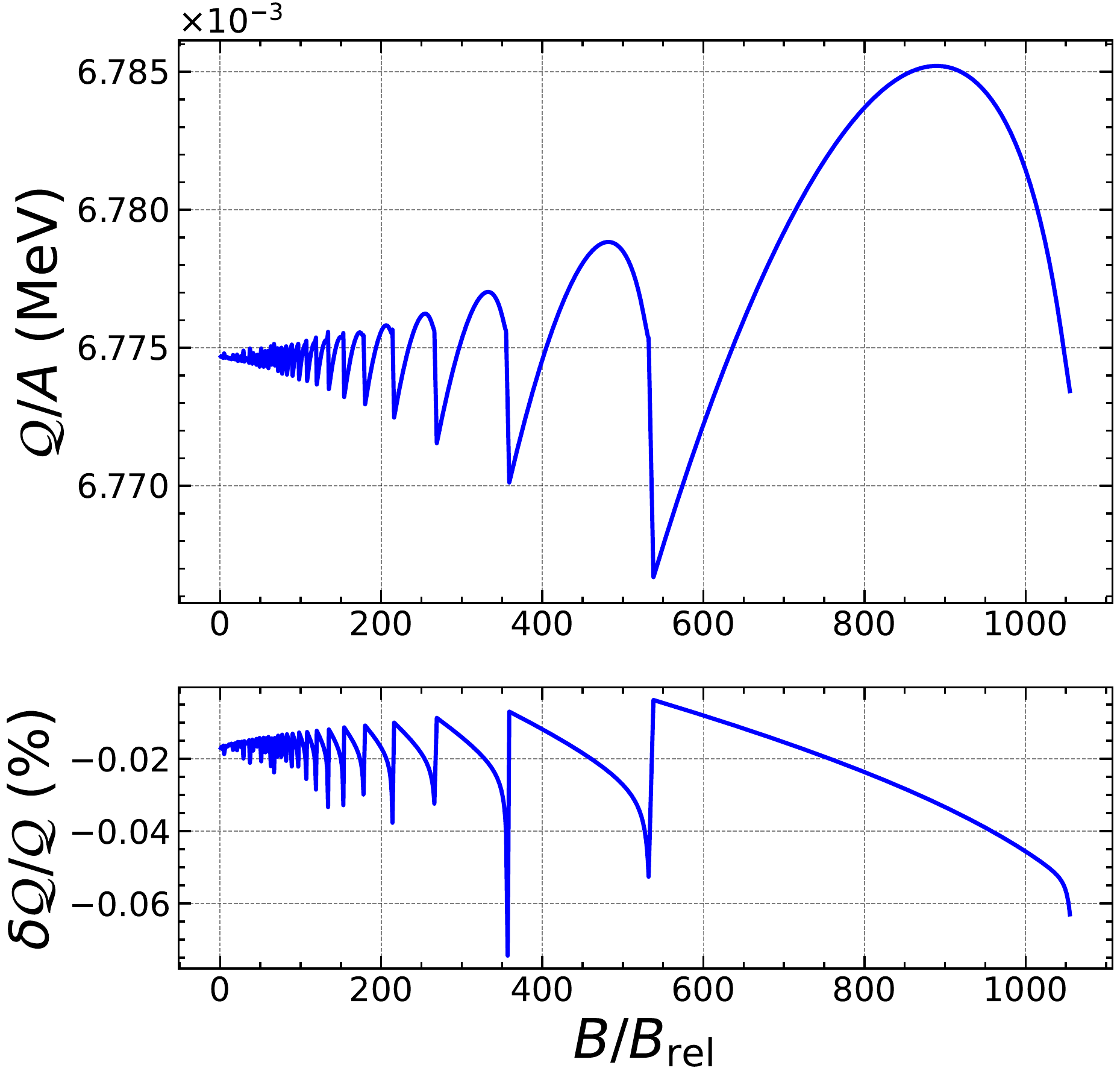}
	\caption{\textbf{Top panel}: Exact heat released per nucleon from electron captures by $^{122}$Zr as a function of the magnetic field strength $B_\star=B/B_{\rm rel}$ up to the onset of the strongly quantizing regime. \textbf{Bottom panel}: Relative error (in \%) of the approximate analytical expression. 
	}
	\label{fig12}
\end{figure}

\begin{figure}[H]
	
	\includegraphics[width=10.5cm]{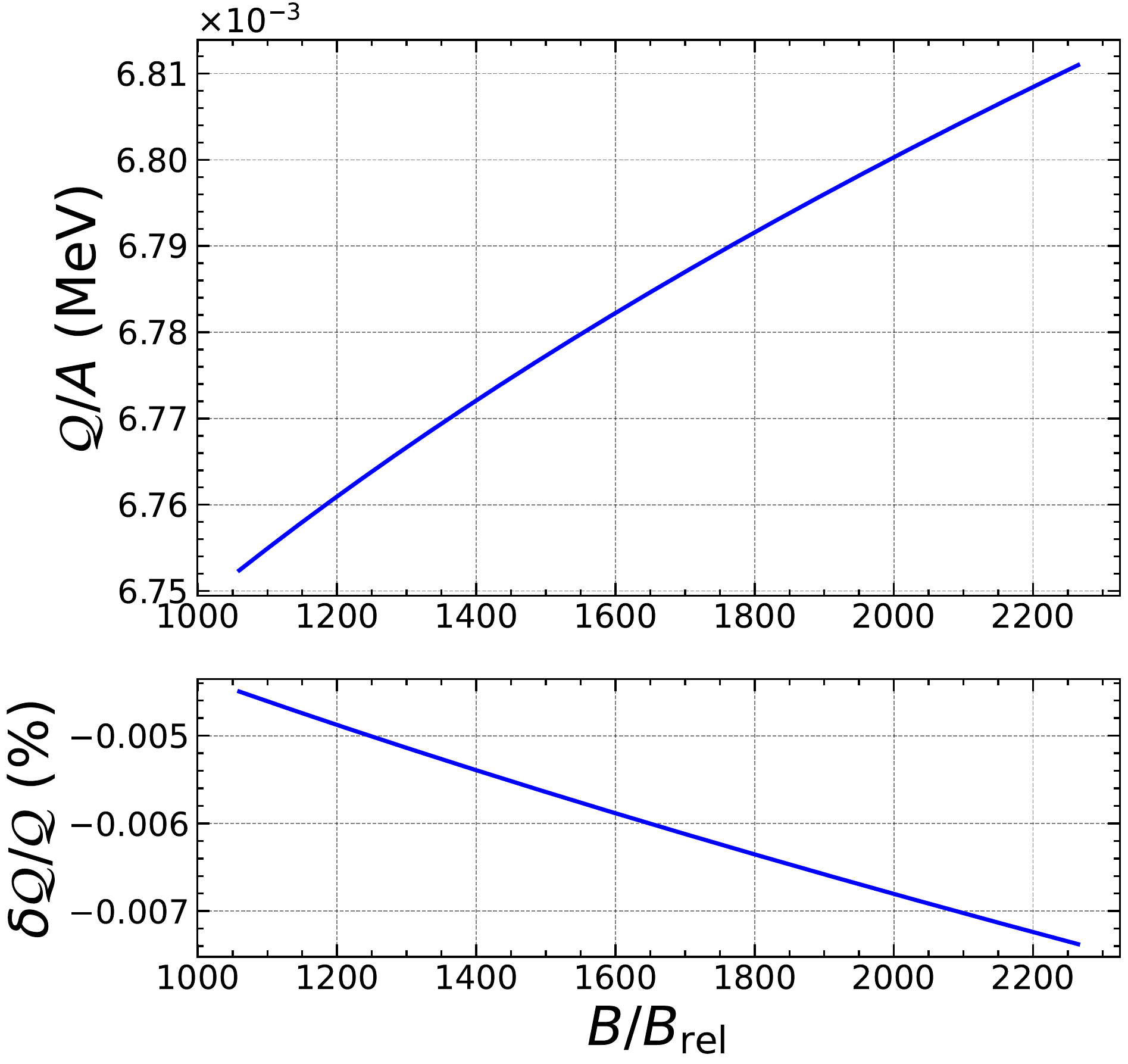}
	\caption{Same as Figure~\ref{fig12} in the strongly quantizing regime. 
	}
	\label{fig13}
\end{figure}

\newpage
To assess the reliability of our analytical treatment, we have numerically solved the exact threshold conditions  $\mu(A,Z,n_e)=\mu(A,Z-1,n_{e1})$ and $P(Z,n_e)=P(Z-1,n_{e1})=P(Z-2,n_{e2})$ without any approximation, i.e.,~using Equations~\eqref{eq:pressure-electron} and \eqref{eq:ne}, to~determine the exact values for the threshold pressure $P_\beta$ and baryon number density $\bar n_\beta$. The~heat deposited is then calculated as $\mathcal{Q}= \mu(A,Z,n_e)-\mu(A,Z-2,n_{e2})$. We have compared these exact results with the approximate analytical formulas.  
As an example, we focus on the electron capture by $^{56}$Fe, considering the ground-state to ground-state transition. 
As shown in Figure~\ref{fig4}, the~quantum oscillations of the threshold density 
are correctly reproduced. The~errors are found to be the largest for specific values of the magnetic field strength  corresponding to exact fillings of Landau--Rabi levels and amount to a few percents, but~drop by about an order of magnitude in the strongly quantizing regime, as shown in Figure~\ref{fig5}. 
As previously discussed in Section~\ref{subsec:weakly-quantizing}, the~expansion in the weakly quantizing regime is 
more reliable for the pressure than for the density. Indeed, the~overall errors
on the threshold pressure are significantly smaller, as~can be observed in 
Figure~\ref{fig2}. In~the strongly quantizing regime, the~errors on $P_\beta$ are of the 
same order as those on $\bar n_\beta$ and are displayed in Figure~\ref{fig3}.  
The heat released, plotted in Figures~\ref{fig6} and \ref{fig7}, also exhibits quantum oscillations, though the 
amplitude is very small. To~a first approximation, the~heat is therefore essentially given by that in the absence 
of magnetic fields, as anticipated. The~analytical formula for $\mathcal{Q}$ is found to be even more precise than formulas 
for the density and pressure. To~check that these error estimates are not specific to the reaction considered, we have also analyzed the electron capture by $^{122}$Zr, which is present in much deeper layers of the outer crust for all the magnetic field strengths. As~can be observed in Figures~\ref{fig8}--\ref{fig13}, the~analytical formulas remain very precise in this other case. We have examined other reactions and reached similar~conclusions.

\vspace{-6pt}

\vspace{-6pt}

\vspace{-6pt}

\vspace{-6pt}

\vspace{-6pt}

\vspace{-6pt}

\section{Conclusions}

We have derived accurate analytical formulas (with typical errors below 1\%) for calculating the threshold density $\bar n_\beta$ and pressure $P_\beta$ for the onset of electron captures by nuclei 
in the shallow layers of magnetar crusts, as~well as the maximum amount of heat released $\mathcal{Q}$ taking into account the Landau--Rabi quantization of electron motion
induced by the magnetic field. We have also obtained formulas for determining the initial constitution of the outer crust. These formulas are applicable over the whole 
range of magnetic fields encountered in neutron stars, from~the weakly quantizing regime to the most extreme situation in which the electrons all lie in the lowest Landau--Rabi~level. 

Using experimental nuclear data supplemented with predictions from the atomic mass model HFB-24, we have calculated all the necessary nuclear parameters to calculate
the shallow heating for any given magnetic field considering both ground-state to ground-state and ground-state to excited-state transitions. Full numerical results 
can be found in~\cite{zenodo22}. Together with the results for the equation of state and superfluid properties published in~\cite{mutafchieva2019,pearson2018,pearson2020,pearson2022,ChamelAllard2021}, they provide consistent microscopic inputs for modelling the magneto-thermal evolution of neutron stars.



\authorcontributions{Conceptualization, N.C.; methodology, N.C.; software, N.C. and A.F.F.; validation, N.C. and A.F.F.; formal analysis, N.C.; investigation, N.C.; writing---original draft preparation, N.C.; writing---review and editing, N.C. and A.F.F.; visualization, N.C.; supervision, N.C.; project administration, N.C. All authors have read and agreed to the published version of the~manuscript. }


\funding{The 
	work of N.C. was funded by Fonds de la Recherche Scientifique-FNRS (Belgium) under Grant Number IISN 4.4502.19. This work was also partially supported by the European Cooperation in Science and Technology Action CA16214 and the CNRS International Research Project (IRP) ``Origine des \'el\'ements lourds dans l’univers: Astres Compacts et Nucl\'eosynth\`ese (ACNu)''. }


\institutionalreview{Not applicable.}

\informedconsent{Not applicable.}

\newpage
\dataavailability{The data analyzed in this paper can be found in the 2016 Atomic Mass Evaluation~\cite{ame2016}, the~BRUSLIB database 
(\url{http://www.astro.ulb.ac.be/bruslib/}, accessed on 9 June 2022, see~\cite{bruslib}), and~the Nuclear Data section of the International Atomic Energy Agency website 
(\url{https://www-nds.iaea.org/relnsd/NdsEnsdf/QueryForm.html}, accessed on 9 June 2022). The~results presented in this study are openly available on the Zenodo repository~\cite{zenodo22}. 
}

\conflictsofinterest{The authors declare no conflict of~interest.}

\begin{adjustwidth}{-\extralength}{0cm}
	
	\reftitle{References}
	
	

\end{adjustwidth}

%


\begin{thebibliography}{999}
		
		\bibitem[{Esposito} \em{et~al.}(2021){Esposito}, {Rea}, and
		{Israel}]{esposito2021}
		{Esposito}, P.; {Rea}, N.; {Israel}, G.L. {Magnetars: A Short Review and Some
			Sparse Considerations}.
		\newblock In {\em Astrophysics and Space Science Library}; {Belloni}, T.M.,
		{M{\'e}ndez}, M., {Zhang}, C., Eds.;  Springer:  Berlin/Heidelberg, Germany, 
		2021; Volume~461, pp.~97--142.
		\newblock {{https://doi.org/10.1007/978-3-662-62110-3\_3}}.
		
		\bibitem[{Duncan} and {Thompson}(1992)]{duncan1992}
		{Duncan}, R.C.; {Thompson}, C.
		\newblock {Formation of Very Strongly Magnetized Neutron Stars: Implications
			for Gamma-Ray Bursts}.
		\newblock {\em \apjl} {\bf 1992}, {\em 392},~L9.
		\newblock {{https://doi.org/10.1086/186413}}.
		
		\bibitem[{Olausen} and {Kaspi}(2014)]{olausen2014}
		{Olausen}, S.A.; {Kaspi}, V.M.
		\newblock {The McGill Magnetar Catalog}.
		\newblock {\em \apjs} {\bf 2014}, {\em 212},~6.
		\newblock {{https://doi.org/10.1088/0067-0049/212/1/6}}.
		
		\bibitem[{Vigan{\`o}} \em{et~al.}(2013){Vigan{\`o}}, {Rea}, {Pons}, {Perna},
		{Aguilera}, and {Miralles}]{vigano2013}
		{Vigan{\`o}}, D.; {Rea}, N.; {Pons}, J.A.; {Perna}, R.; {Aguilera}, D.N.;
		{Miralles}, J.A.
		\newblock {Unifying the observational diversity of isolated neutron stars via
			magneto-thermal evolution models}.
		\newblock {\em \mnras} {\bf 2013}, {\em 434},~123--141.
		\newblock {{https://doi.org/10.1093/mnras/stt1008}}.
		
		\bibitem[{Beloborodov} and {Li}(2016)]{beloborodov2016}
		{Beloborodov}, A.M.; {Li}, X.
		\newblock {Magnetar Heating}.
		\newblock {\em \apj} {\bf 2016}, {\em 833},~261.
		\newblock {{https://doi.org/10.3847/1538-4357/833/2/261}}.
		
		\bibitem[{De Grandis} \em{et~al.}(2020){De Grandis}, {Turolla}, {Wood}, {Zane},
		{Taverna}, and {Gourgouliatos}]{degrandis2020}
		{De Grandis}, D.; {Turolla}, R.; {Wood}, T.S.; {Zane}, S.; {Taverna}, R.;
		{Gourgouliatos}, K.N.
		\newblock {Three-dimensional Modeling of the Magnetothermal Evolution of
			Neutron Stars: Method and Test Cases}.
		\newblock {\em \apj} {\bf 2020}, {\em 903},~40.
		\newblock {{https://doi.org/10.3847/1538-4357/abb6f9}}.
		
		\bibitem[{Fantina} \em{et~al.}(2020){Fantina}, {De Ridder}, {Chamel}, and
		{Gulminelli}]{fantina2020}
		{Fantina}, A.F.; {De Ridder}, S.; {Chamel}, N.; {Gulminelli}, F.
		\newblock {Crystallization of the outer crust of a non-accreting neutron star}.
		\newblock {\em \aap} {\bf 2020}, {\em 633},~A149.
		\newblock {{https://doi.org/10.1051/0004-6361/201936359}}.
		
		\bibitem[{Carreau} \em{et~al.}(2020){Carreau}, {Gulminelli}, {Chamel},
		{Fantina}, and {Pearson}]{carreau2020}
		{Carreau}, T.; {Gulminelli}, F.; {Chamel}, N.; {Fantina}, A.F.; {Pearson}, J.M.
		\newblock {Crystallization of the inner crust of a neutron star and the
			influence of shell effects}.
		\newblock {\em \aap} {\bf 2020}, {\em 635},~A84.
		\newblock {{https://doi.org/10.1051/0004-6361/201937236}}.
		
		\bibitem[{Kaminker} \em{et~al.}(2006){Kaminker}, {Yakovlev}, {Potekhin},
		{Shibazaki}, {Shternin}, and {Gnedin}]{kaminker2006}
		{Kaminker}, A.D.; {Yakovlev}, D.G.; {Potekhin}, A.Y.; {Shibazaki}, N.;
		{Shternin}, P.S.; {Gnedin}, O.Y.
		\newblock {Magnetars as cooling neutron stars with internal heating}.
		\newblock {\em \mnras} {\bf 2006}, {\em 371},~477--483.
		\newblock {{https://doi.org/10.1111/j.1365-2966.2006.10680.x}}.
		
		\bibitem[{Kaminker} \em{et~al.}(2009){Kaminker}, {Potekhin}, {Yakovlev}, and
		{Chabrier}]{kaminker2009}
		{Kaminker}, A.D.; {Potekhin}, A.Y.; {Yakovlev}, D.G.; {Chabrier}, G.
		\newblock {Heating and cooling of magnetars with accreted envelopes}.
		\newblock {\em \mnras} {\bf 2009}, {\em 395},~2257--2267.
		\newblock {{https://doi.org/10.1111/j.1365-2966.2009.14693.x}}.
		
		\bibitem[{Cooper} and {Kaplan}(2010)]{cooper2010}
		{Cooper}, R.L.; {Kaplan}, D.L.
		\newblock {Magnetic Field-Decay-Induced Electron Captures: A Strong Heat Source
			in Magnetar Crusts}.
		\newblock {\em \apjl} {\bf 2010}, {\em 708},~L80--L83.
		\newblock {{https://doi.org/10.1088/2041-8205/708/2/L80}}.
		
		\bibitem[{Haensel} and {Zdunik}(1990)]{haensel1990}
		{Haensel}, P.; {Zdunik}, J.L.
		\newblock {Non-equilibrium processes in the crust of an accreting neutron
			star}.
		\newblock {\em \aap} {\bf 1990}, {\em 227},~431--436.
		
		\bibitem[{Chamel} \em{et~al.}(2021){Chamel}, {Fantina}, {Suleiman}, {Zdunik},
		and {Haensel}]{chamel2021}
		{Chamel}, N.; {Fantina}, A.F.; {Suleiman}, L.; {Zdunik}, J.L.; {Haensel}, P.
		\newblock {Heating in Magnetar Crusts from Electron Captures}.
		\newblock {\em Universe} {\bf 2021}, {\em 7},~193.
		\newblock {{https://doi.org/10.3390/universe7060193}}.
		
		\bibitem[{Goriely} \em{et~al.}(2013){Goriely}, {Chamel}, and
		{Pearson}]{goriely2013}
		{Goriely}, S.; {Chamel}, N.; {Pearson}, J.M.
		\newblock {Further explorations of Skyrme-Hartree-Fock-Bogoliubov mass
			formulas. XIII. The 2012 atomic mass evaluation and the symmetry
			coefficient}.
		\newblock {\em \prc} {\bf 2013}, {\em 88},~024308.
		\newblock {{https://doi.org/10.1103/PhysRevC.88.024308}}.
		
		\bibitem[{Pearson} \em{et~al.}(2018){Pearson}, {Chamel}, {Potekhin}, {Fantina},
		{Ducoin}, {Dutta}, and {Goriely}]{pearson2018}
		{Pearson}, J.M.; {Chamel}, N.; {Potekhin}, A.Y.; {Fantina}, A.F.; {Ducoin}, C.;
		{Dutta}, A.K.; {Goriely}, S.
		\newblock {Unified equations of state for cold non-accreting neutron stars with
			Brussels-Montreal functionals---I. Role of symmetry energy}.
		\newblock {\em \mnras} {\bf 2018}, {\em 481},~2994--3026.
		\newblock {{https://doi.org/10.1093/mnras/sty2413}}.
		
		\bibitem[{Pearson} \em{et~al.}(2020){Pearson}, {Chamel}, and
		{Potekhin}]{pearson2020}
		\textls[-20]{{Pearson}, J.M.; {Chamel}, N.; {Potekhin}, A.Y.}
		\newblock \textls[-20]{{Unified equations of state for cold nonaccreting neutron stars with
				Brussels-Montreal functionals. II. Pasta phases in semiclassical
				approximation}.}
		\newblock {\em \prc} {\bf 2020}, {\em 101},~015802.
		\newblock {{https://doi.org/10.1103/PhysRevC.101.015802}}.
		
		\bibitem[{Pearson} and {Chamel}(2022)]{pearson2022}
		{Pearson}, J.M.; {Chamel}, N.
		\newblock {Unified equations of state for cold nonaccreting neutron stars with
			Brussels-Montreal functionals. III. Inclusion of microscopic corrections to
			pasta phases}.
		\newblock {\em \prc} {\bf 2022}, {\em 105},~015803.
		\newblock {{https://doi.org/10.1103/PhysRevC.105.015803}}.
		
		\bibitem[{Mutafchieva} \em{et~al.}(2019){Mutafchieva}, {Chamel}, {Stoyanov},
		{Pearson}, and {Mihailov}]{mutafchieva2019}
		{Mutafchieva}, Y.D.; {Chamel}, N.; {Stoyanov}, Z.K.; {Pearson}, J.M.;
		{Mihailov}, L.M.
		\newblock {Role of Landau-Rabi quantization of electron motion on the crust of
			magnetars within the nuclear energy density functional theory}.
		\newblock {\em \prc} {\bf 2019}, {\em 99},~055805.
		\newblock {{https://doi.org/10.1103/PhysRevC.99.055805}}.
		
		\bibitem[{Rabi}(1928)]{rabi1928}
		{Rabi}, I.I.
		\newblock {Das freie Elektron im homogenen Magnetfeld nach der Diracschen
			Theorie}.
		\newblock {\em Z. Fur. Phys.} {\bf 1928}, {\em 49},~507--511.
		\newblock {{https://doi.org/10.1007/BF01333634}}.
		
		\bibitem[{Landau}(1930)]{landau1930}
		{Landau}, L.
		\newblock {Diamagnetismus der Metalle}.
		\newblock {\em Z. Fur Phys.} {\bf 1930}, {\em 64},~629--637.
		\newblock {{https://doi.org/10.1007/BF01397213}}.
		
		\bibitem[{Tiengo} \em{et~al.}(2013){Tiengo}, {Esposito}, {Mereghetti},
		{Turolla}, {Nobili}, {Gastaldello}, {G{\"o}tz}, {Israel}, {Rea}, {Stella},
		{Zane}, and {Bignami}]{tiengo2013}
		{Tiengo}, A.; {Esposito}, P.; {Mereghetti}, S.; {Turolla}, R.; {Nobili}, L.;
		{Gastaldello}, F.; {G{\"o}tz}, D.; {Israel}, G.L.; {Rea}, N.; {Stella}, L.;
		et~al.
		\newblock {A variable absorption feature in the X-ray spectrum of a magnetar}.
		\newblock {\em \nat} {\bf 2013}, {\em 500},~312--314.
		\newblock {{https://doi.org/10.1038/nature12386}}.
		
		\bibitem[{An} \em{et~al.}(2014){An}, {Kaspi}, {Beloborodov}, {Kouveliotou},
		{Archibald}, {Boggs}, {Christensen}, {Craig}, {Gotthelf}, {Grefenstette},
		{Hailey}, {Harrison}, {Madsen}, {Mori}, {Stern}, and {Zhang}]{hongjun2014}
		{An}, H.; {Kaspi}, V.M.; {Beloborodov}, A.M.; {Kouveliotou}, C.; {Archibald},
		R.F.; {Boggs}, S.E.; {Christensen}, F.E.; {Craig}, W.W.; {Gotthelf}, E.V.;
		{Grefenstette}, B.W.;  et~al.
		\newblock {NuSTAR Observations of X-Ray Bursts from the Magnetar 1E
			1048.1-5937}.
		\newblock {\em \apj} {\bf 2014}, {\em 790},~60.
		\newblock {{https://doi.org/10.1088/0004-637X/790/1/60}}.
		
		\bibitem[{Ury{\={u}}} \em{et~al.}(2019){Ury{\={u}}}, {Yoshida}, {Gourgoulhon},
		{Markakis}, {Fujisawa}, {Tsokaros}, {Taniguchi}, and {Eriguchi}]{uryu2019}
		{Ury{\={u}}}, K.; {Yoshida}, S.; {Gourgoulhon}, E.; {Markakis}, C.; {Fujisawa},
		K.; {Tsokaros}, A.; {Taniguchi}, K.; {Eriguchi}, Y.
		\newblock {New code for equilibriums and quasiequilibrium initial data of
			compact objects. IV. Rotating relativistic stars with mixed poloidal and
			toroidal magnetic fields}.
		\newblock {\em \prd} {\bf 2019}, {\em 100},~123019.
		\newblock {{https://doi.org/10.1103/PhysRevD.100.123019}}.
		
		\bibitem[{Chamel} \em{et~al.}(2015){Chamel}, {Fantina}, {Zdunik}, and
		{Haensel}]{chamel2015b}
		{Chamel}, N.; {Fantina}, A.F.; {Zdunik}, J.L.; {Haensel}, P.
		\newblock {Neutron drip transition in accreting and nonaccreting neutron star
			crusts}.
		\newblock {\em \prc} {\bf 2015}, {\em 91},~055803.
		\newblock {{https://doi.org/10.1103/PhysRevC.91.055803}}.
		
		\bibitem[Haensel \em{et~al.}(2007)Haensel, Potekhin, and Yakovlev]{haensel2007}
		Haensel, P.; Potekhin, A.Y.; Yakovlev, D.G.
		\newblock {\em {Neutron Stars. 1. Equation of State and Structure}}; Springer:
		New York, NY, USA, 2007.
		
		\bibitem[{Van Vleck}(1932)]{vanvleck1932}
		{Van Vleck}, J.H.
		\newblock {\em The Theory of Electric and Magnetic Susceptibilities}; Oxford
		University Press: London, UK, 1932.
		
		\bibitem[{Baiko}(2009)]{baiko2009}
		{Baiko}, D.A.
		\newblock {Coulomb crystals in the magnetic field}.
		\newblock {\em \pre} {\bf 2009}, {\em 80},~046405.
		\newblock {{https://doi.org/10.1103/PhysRevE.80.046405}}.
		
		\bibitem[{Baiko} \em{et~al.}(2001){Baiko}, {Potekhin}, and
		{Yakovlev}]{baiko2001}
		{Baiko}, D.A.; {Potekhin}, A.Y.; {Yakovlev}, D.G.
		\newblock {Thermodynamic functions of harmonic Coulomb crystals}.
		\newblock {\em \pre} {\bf 2001}, {\em 64},~057402.
		\newblock {{https://doi.org/10.1103/PhysRevE.64.057402}}.
		
		\bibitem[{Potekhin} and {Chabrier}(2013)]{potekhin2013}
		{Potekhin}, A.Y.; {Chabrier}, G.
		\newblock {Equation of state for magnetized Coulomb plasmas}.
		\newblock {\em \aap} {\bf 2013}, {\em 550},~A43.
		\newblock {{https://doi.org/10.1051/0004-6361/201220082}}.
		
		\bibitem[{Salpeter}(1954)]{salpeter1954}
		{Salpeter}, E.E.
		\newblock {Electrons Screening and Thermonuclear Reactions}.
		\newblock {\em Aust. J. Phys.} {\bf 1954}, {\em 7},~373.
		\newblock {{https://doi.org/10.1071/PH540373}}.
		
		\bibitem[{Dib} and {Espinosa}(2001)]{dib2001}
		{Dib}, C.O.; {Espinosa}, O.
		\newblock {The magnetized electron gas in terms of Hurwitz zeta functions}.
		\newblock {\em Nucl. Phys. B} {\bf 2001}, {\em 612},~492--518.
		\newblock {{https://doi.org/10.1016/S0550-3213(01)00360-1}}.
		
		\bibitem[{Chamel} and {Fantina}(2015)]{fantina2015}
		{Chamel}, N.; {Fantina}, A.F.
		\newblock {Electron capture instability in magnetic and nonmagnetic white
			dwarfs}.
		\newblock {\em \prd} {\bf 2015}, {\em 92},~023008.
		\newblock {{https://doi.org/10.1103/PhysRevD.92.023008}}.
		
		\bibitem[{Chamel}(2020)]{chamel2020}
		{Chamel}, N.
		\newblock {Analytical determination of the structure of the outer crust of a
			cold nonaccreted neutron star}.
		\newblock {\em \prc} {\bf 2020}, {\em 101},~032801.
		\newblock {{https://doi.org/10.1103/PhysRevC.101.032801}}.
		
		\bibitem[{Chamel} and {Stoyanov}(2020)]{chamelstoyanov2020}
		{Chamel}, N.; {Stoyanov}, Zh.K.
		\newblock {Analytical determination of the structure of the outer crust of a
			cold nonaccreted neutron star: Extension to strongly quantizing magnetic
			fields}.
		\newblock {\em \prc} {\bf 2020}, {\em 101},~065802.
		\newblock {{https://doi.org/10.1103/PhysRevC.101.065802}}.
		
		\bibitem[{Chamel}(2020)]{chamel2020zndo}
		{Chamel}, N.
		\newblock {\emph{Equilibrium Structure of the Outer Crust of a Cold Nonaccreted
				Neutron Star}}; Zenodo: Geneva, Switzerland, 2020.
		\newblock {{https://doi.org/10.5281/zenodo.3719439}}.
		
		\bibitem[{Chamel} and {Stoyanov}(2020)]{chamelstoyanov2020zndo}
		{Chamel}, N.; {Stoyanov}, Zh.
		\newblock {\emph{Equilibrium Structure of the Outer Crust of a Magnetar}}; Zenodo: Geneva, Switzerland, 2020.
		\newblock {{https://doi.org/10.5281/zenodo.3839787}}.
		
		\bibitem[{Pons} and {Vigan{\`o}}(2019)]{pons2019}
		{Pons}, J.A.; {Vigan{\`o}}, D.
		\newblock {Magnetic, thermal and rotational evolution of isolated neutron
			stars}.
		\newblock {\em Living Rev. Comput. Astrophys.} {\bf 2019}, {\em
			5},~3.
		\newblock {{https://doi.org/10.1007/s41115-019-0006-7}}.
		
		\bibitem[Chamel and Fantina(2016)]{chamelfantina2016}
		Chamel, N.; Fantina, A.F.
		\newblock Binary and ternary ionic compounds in the outer crust of a cold
		nonaccreting neutron star.
		\newblock {\em Phys. Rev. C} {\bf 2016}, {\em 94},~065802.
		\newblock {{https://doi.org/10.1103/PhysRevC.94.065802}}.
		
		\bibitem[{Pe{\~n}a Arteaga} \em{et~al.}(2011){Pe{\~n}a Arteaga}, {Grasso},
		{Khan}, and {Ring}]{arteaga2011}
		{Pe{\~n}a Arteaga}, D.; {Grasso}, M.; {Khan}, E.; {Ring}, P.
		\newblock {Nuclear structure in strong magnetic fields: Nuclei in the crust of
			a magnetar}.
		\newblock {\em \prc} {\bf 2011}, {\em 84},~045806.
		\newblock {{https://doi.org/10.1103/PhysRevC.84.045806}}.
		
		\bibitem[{Fantina} \em{et~al.}(2016){Fantina}, {Chamel}, {Mutafchieva},
		{Stoyanov}, {Mihailov}, and {Pavlov}]{fantina2016a}
		{Fantina}, A.F.; {Chamel}, N.; {Mutafchieva}, Y.D.; {Stoyanov}, Zh.K.;
		{Mihailov}, L.M.; {Pavlov}, R.L.
		\newblock {Role of the symmetry energy on the neutron-drip transition in
			accreting and nonaccreting neutron stars}.
		\newblock {\em \prc} {\bf 2016}, {\em 93},~015801.
		\newblock {{https://doi.org/10.1103/PhysRevC.93.015801}}.
		
		\bibitem[{Wang} \em{et~al.}(2017){Wang}, {Audi}, {Kondev}, {Huang}, {Naimi},
		and {Xu}]{ame2016}
		{Wang}, M.; {Audi}, G.; {Kondev}, F.G.; {Huang}, W.J.; {Naimi}, S.; {Xu}, X.
		\newblock {The AME2016 atomic mass evaluation (II). Tables, graphs and
			references}.
		\newblock {\em Chin. Phys. C} {\bf 2017}, {\em 41},~030003.
		\newblock {{https://doi.org/10.1088/1674-1137/41/3/030003}}.
		
		\bibitem[{Xu} \em{et~al.}(2013){Xu}, {Goriely}, {Jorissen}, {Chen}, and
		{Arnould}]{bruslib}
		{Xu}, Y.; {Goriely}, S.; {Jorissen}, A.; {Chen}, G.L.; {Arnould}, M.
		\newblock {Databases and tools for nuclear astrophysics applications. BRUSsels
			Nuclear LIBrary (BRUSLIB), Nuclear Astrophysics Compilation of REactions II
			(NACRE II) and Nuclear NETwork GENerator (NETGEN)}.
		\newblock {\em \aap} {\bf 2013}, {\em 549},~A106.
		\newblock {{https://doi.org/10.1051/0004-6361/201220537}}.
		
		\bibitem[Allard and Chamel(2021)]{ChamelAllard2021}
		Allard, V.; Chamel, N.
		\newblock 1S0 Pairing Gaps, Chemical Potentials and Entrainment Matrix in
		Superfluid Neutron-Star Cores for the Brussels-Montreal Functionals.
		\newblock {\em Universe} {\bf 2021}, {\em 7},~470.
		\newblock {{https://doi.org/10.3390/universe7120470}}.
		
		\bibitem[{Perot} \em{et~al.}(2019){Perot}, {Chamel}, and {Sourie}]{perot2019}
		{Perot}, L.; {Chamel}, N.; {Sourie}, A.
		\newblock {Role of the symmetry energy and the neutron-matter stiffness on the
			tidal deformability of a neutron star with unified equations of state}.
		\newblock {\em \prc} {\bf 2019}, {\em 100},~035801.
		\newblock {{https://doi.org/10.1103/PhysRevC.100.035801}}.
		
		\bibitem[Chamel and Fantina(2022)]{zenodo22}
		Chamel, N.; Fantina, A.F.
		\newblock {\emph{Onset of Electron Captures and Shallow Heating in Magnetars [Data
				Set]}}; Zenodo: Geneva, Switzerland, 2022.
		\newblock {{https://doi.org/10.5281/zenodo.6604639}}.
		
	\end{thebibliography}
\end{document}